%% file: main.tex
\documentclass[letterpaper,twocolumn,10pt]{article}
\usepackage{usenix2019_v3}

\usepackage[utf8]{inputenc}

\usepackage{xspace}
\usepackage{xcolor}
\usepackage{paralist}
\usepackage{wrapfig}
\usepackage{multirow}
\usepackage{listings}
\usepackage{makecell}
\usepackage{boxedminipage}
\usepackage{booktabs}
\usepackage{tabularx}
\usepackage{url}
\usepackage{amsmath}
\usepackage{gensymb}

\usepackage{breakurl} 
\usepackage{boldline}
\usepackage{tikz}
\usepackage{amsmath}
\usepackage{hyperref}
\usepackage{pdfpages}

\usepackage{multirow}
\usepackage{subcaption}
\usepackage{caption}
\usepackage{graphicx}
\graphicspath{ {./figs/} }
\usepackage[ruled,vlined,linesnumbered]{algorithm2e}
\usepackage{booktabs}

\usepackage{enumitem}

\begin{document}

\date{}

\title{Penetration Vision through Virtual Reality Headsets: \\ Identifying 360-degree Videos from Head Movements}

\author{
{\rm Anh Nguyen}\\
George Mason University
\and
{\rm Xiaokuan Zhang}\\
George Mason University
\and
{\rm Zhisheng Yan}\\
George Mason University
} 
\input{macro}
\maketitle

\begin{abstract}
In this paper, we present the \textit{first} contactless side-channel attack 
for identifying 360\degrees videos being viewed in a Virtual Reality (VR) Head Mounted Display (HMD).
Although the video content is displayed inside the HMD without any external exposure, we observe that user head movements are driven by the video content, which creates a unique side channel that does not exist in traditional 2D videos.
By recording the user whose vision is blocked by the HMD via a malicious camera, an attacker can analyze the correlation between the user's head movements and the victim video to infer the video title.

To exploit this new vulnerability, we present \sysname, a system for identifying 360\degrees videos from recordings of user head movements. 
\sysname is empowered by an HMD-based head movement estimation scheme to extract a head movement trace from the recording and a video saliency-based trace-fingerprint matching framework to
infer the video title. 
Evaluation results show that 
\sysname achieves over 96\% of accuracy for video identification and is robust under different recording environments. 
Moreover, \sysname maintains its effectiveness in the open-world identification scenario.
\end{abstract}

\input{1-intro-new}

\section{Background and Motivation}\label{sec_mot}

\subsection{360\degree~Videos and Head Movement}
360\degree\ video viewing is an important application in VR. To view a 360\degree\ video using an HMD, a user must first select the video of interest via the VR hand controller. Once the video playback starts, the VR engine of the video player decodes the video into a spherical scene wrapping around the user's head. A subset of the sphere to which the head is oriented, i.e., the viewport, is rendered and displayed on the HMD screen. This way, the user can simply move his/her head around to explore the 360\degree\ video in an immersive hands-free way.\looseness=-1

It is well-known that the displayed 360\degree\ video drives the head movement of the user \cite{nguyen2018your,appcfan}. This is because the human vision system tends to selectively pay attention to some salient objects or events in a visual scene instead of processing every single pixel. As a result, the user follows his/her visual attention and moves his/her head around when viewing the 360\degree\ video. The user may explore and glance through the content or fixate on a region of interest. Low-level content features of 360\degree\ videos, such as color, contrast, and textures, can affect visual attention and head movement. High-level content features such as human faces and notable objects are also critical to the head movement. 

More importantly, such a correlation between the 360\degree\ video and the head movement are similar across users \cite{dscorbillon,dslo,nguyen2019saliency,carlsson2021cross}. In other words, different users would present similar head movement patterns when viewing the same video. Therefore, even though we have no knowledge about the victim, we can collect fingerprints for videos of interest and utilize this correlation to match the victim's head movement trace with all the fingerprints. By finding the most similar fingerprint, we can infer the video being viewed.

\subsection{Video Fingerprint and Saliency Maps}
\label{sec_mot_how}
\begin{figure}[t]
\centering
\begin{tabular}{cc}
\subfloat{\includegraphics[width=0.19\textwidth]{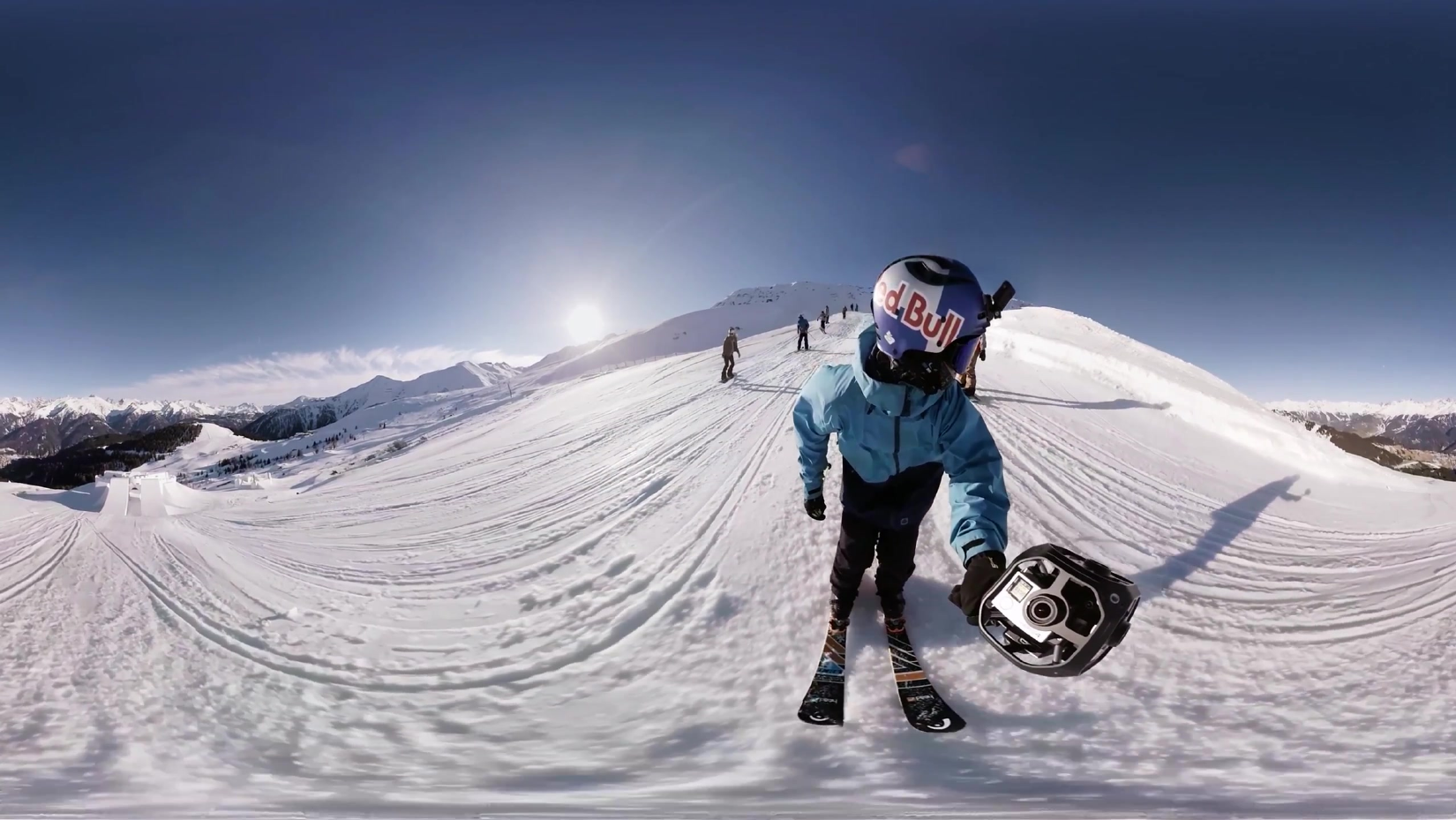}
\label{fig_motivation_image}}&

\subfloat{\includegraphics[width=0.19\textwidth]{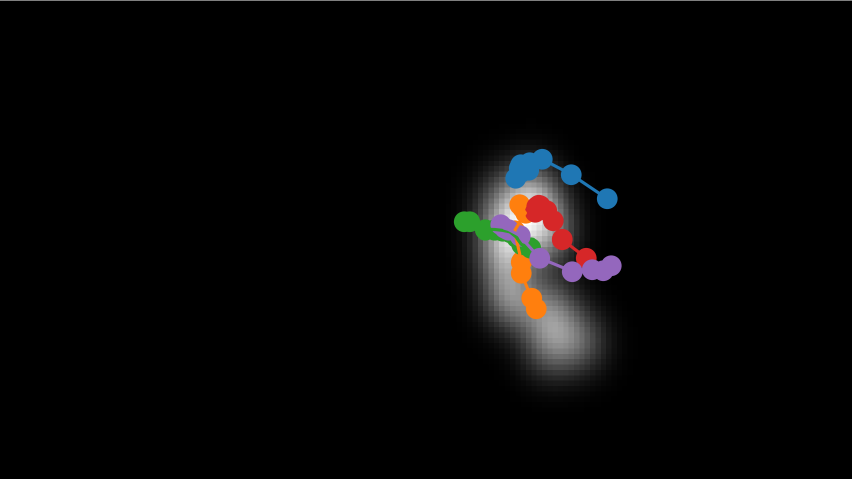}
\label{fig_motivation_saliencymov}}\\
\end{tabular}
\caption{The head movement traces of different users (right figure, 4 colors) on the same video may be different in the time domain, but they do present a similar head movement pattern on regions of interest (right figure, brighter areas).
}
\label{fig_motivation}
\end{figure}

Despite the intuitive principle, matching head movements with videos is challenging. Our study discovers that even though similar patterns exist in different head movement traces of the same video, it does not mean that these head movement time series are exactly the same. In fact, these head movement traces of the same video may differ from each other significantly in the time domain due to the time-varying physical and emotional states of the viewer and external factors of the viewing environment \cite{katsuki2014bottom,lookAround17}. For example, two viewers may look at the same object at different moments or view different parts of the same object at the same time. \autoref{fig_motivation} illustrates this fact via one-second head movement traces of four users (right) viewing a sample 360\degree\ video (left). We can see that different users' head movement traces (marked by four different colors) deviate from each other. Therefore, because of the randomness of head movement in the time domain, even if we can obtain the head movement trace for a video, it is infeasible to use it as the video fingerprint and directly match it with the victim's head movement.

Fortunately, we also observe that different users' head movements tend to focus on the same region, e.g., the bright area in \autoref{fig_motivation} (right). This is attributed to the fact that this region contains attentive content, i.e., the blueish upper body of the human, and thus drives all users' heads to move here. This phenomenon is consistent with previous findings in viewing behavior of 360\degrees\ videos \cite{carlsson2021cross,nguyen2018your,appcfan}. In essence, these salient regions of a 360\degree\ video are highly correlated to different users' head movement traces. They actually represent the similar head movement pattern across different users' traces. Hence, we propose to fingerprint 360\degree\ videos by their salient regions and then match the victim's head movement trace with the fingerprint (salient regions). 

\begin{figure}[!t]
\centering
\subfloat{\includegraphics[width=0.23\columnwidth]{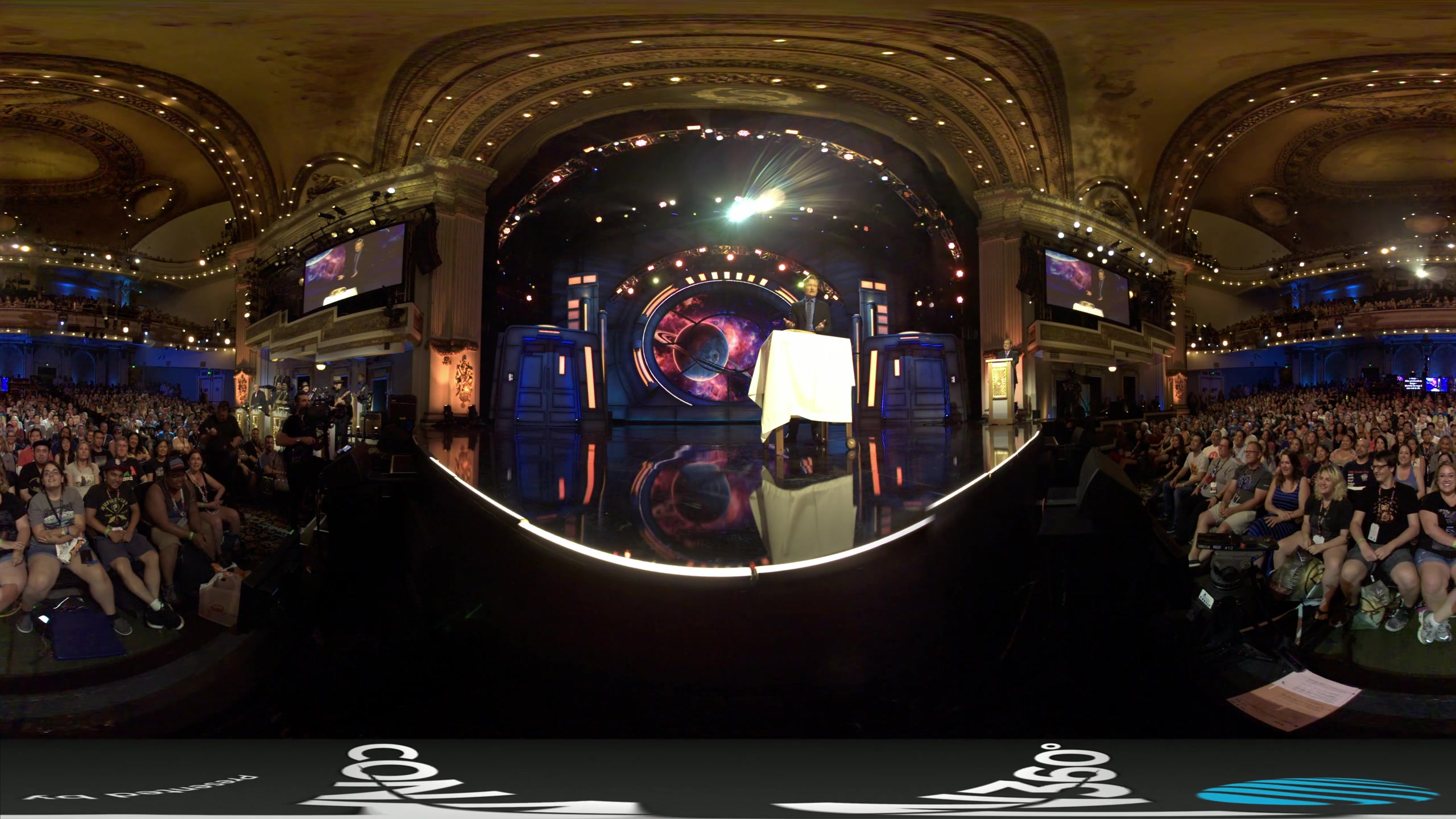}}
\hspace{0.01\columnwidth}
\subfloat{\includegraphics[width=0.23\columnwidth]{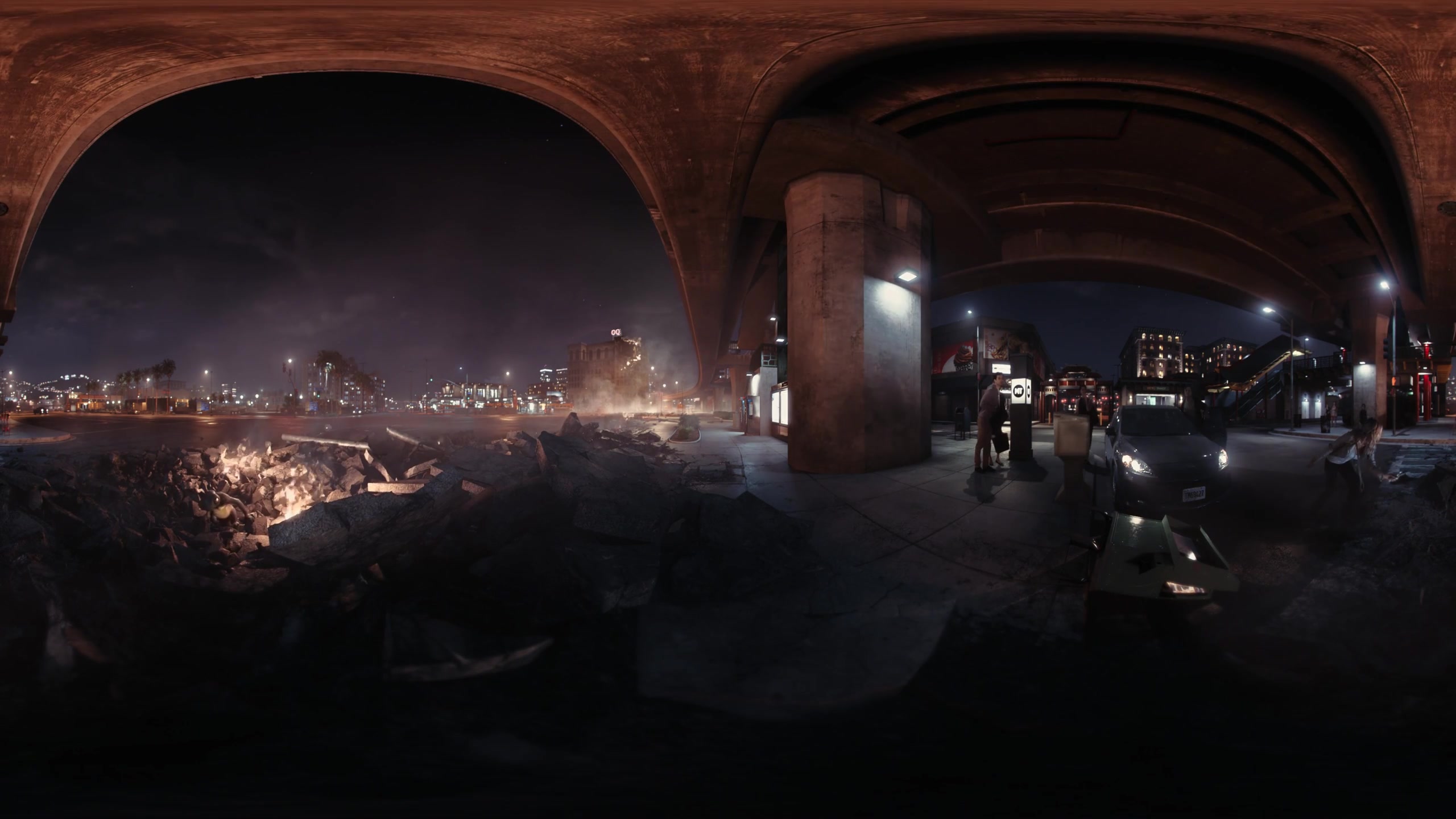}}
\hspace{0.01\columnwidth}
\subfloat{\includegraphics[width=0.23\columnwidth]{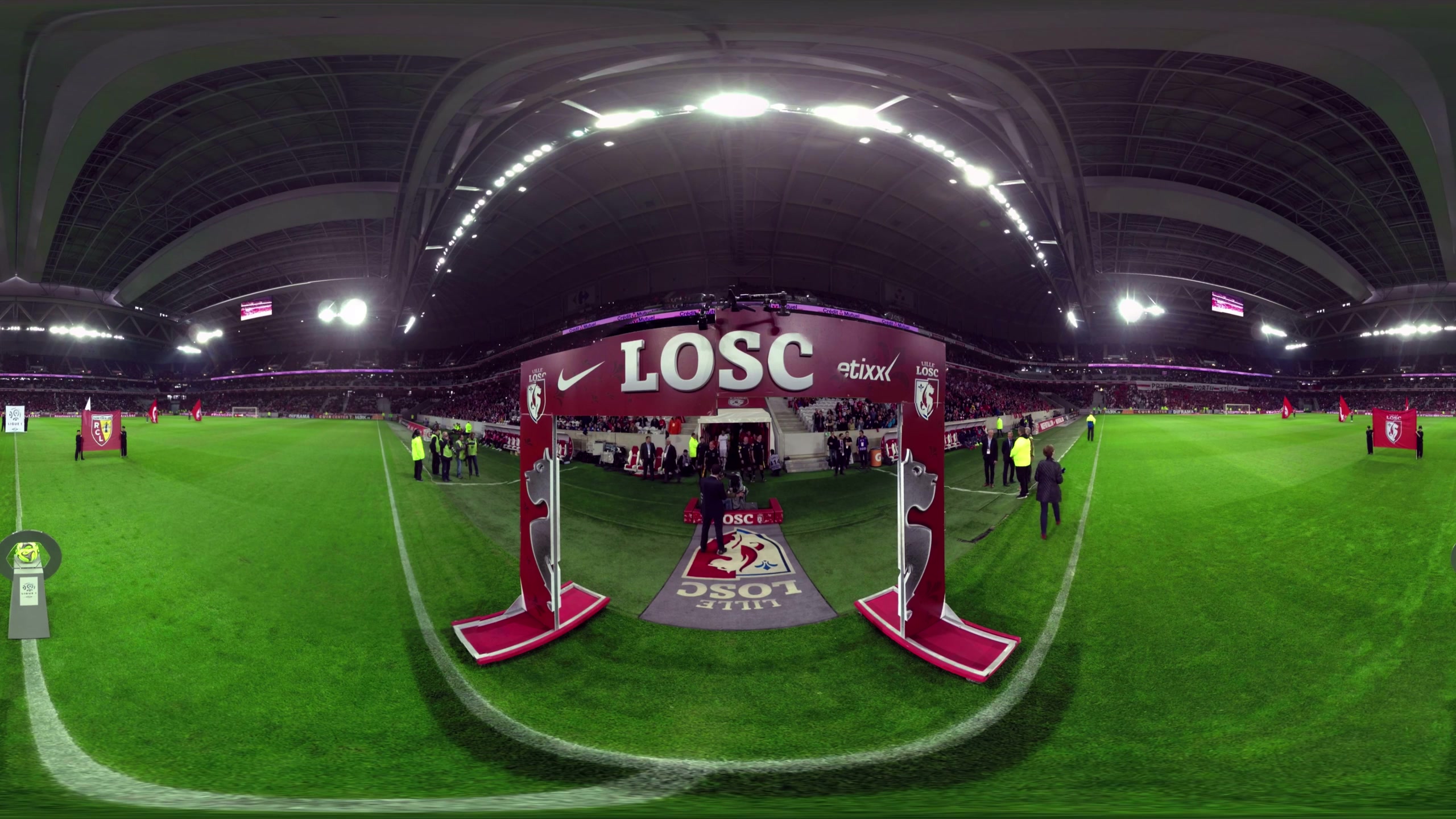}}
\hspace{0.01\columnwidth}
\subfloat{\includegraphics[width=0.23\columnwidth]{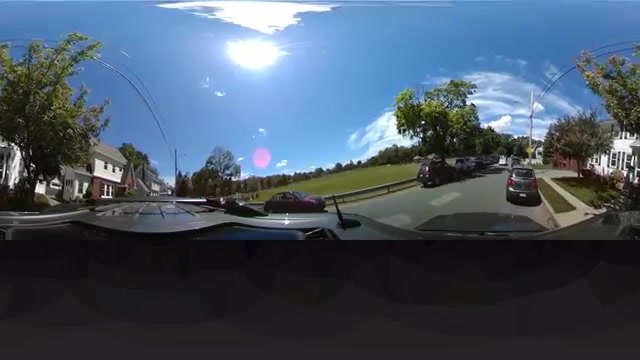}}\\[0ex]
\subfloat{\includegraphics[width=0.23\columnwidth]{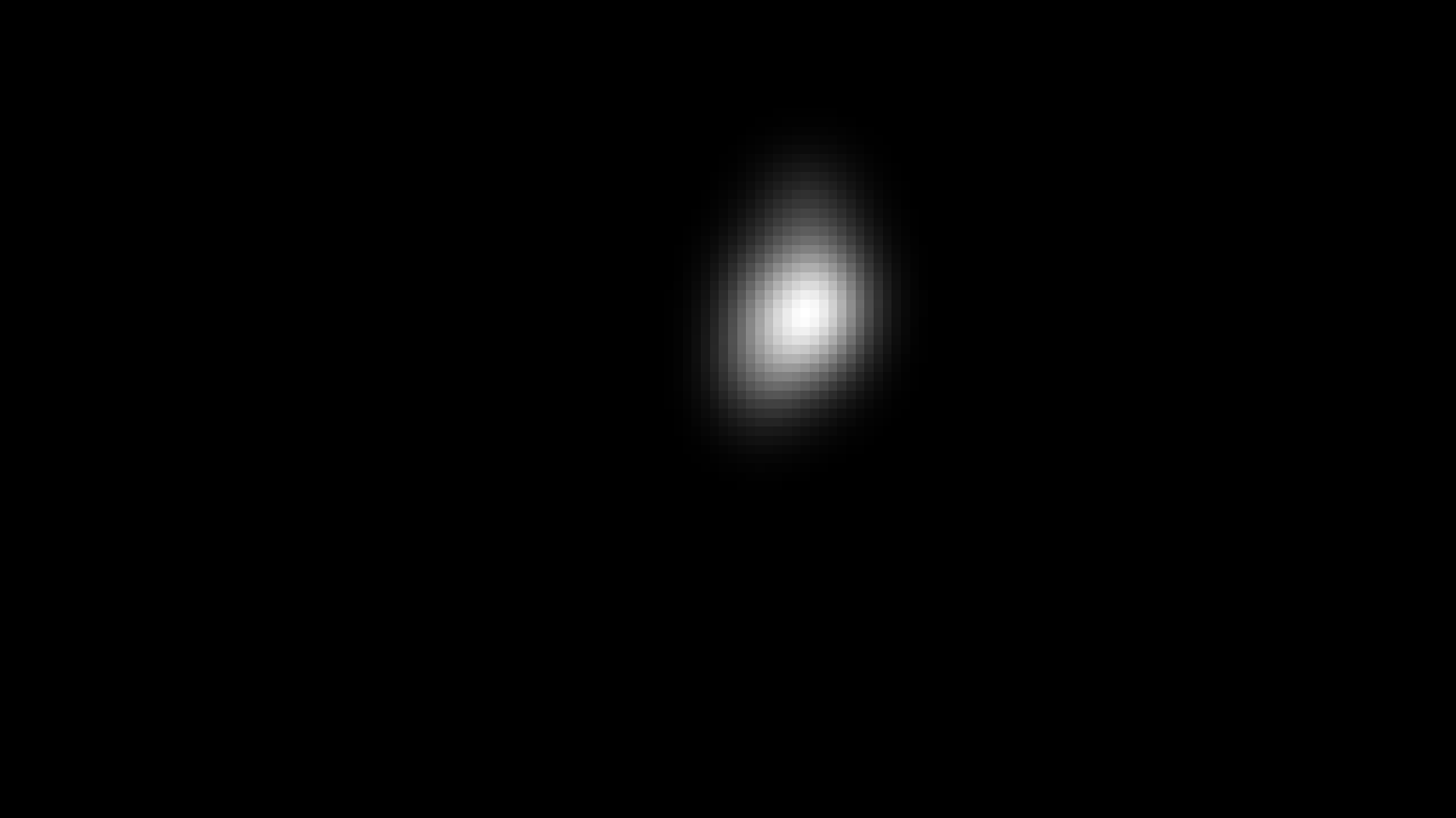}}
\hspace{0.01\columnwidth}
\subfloat{\includegraphics[width=0.23\columnwidth]{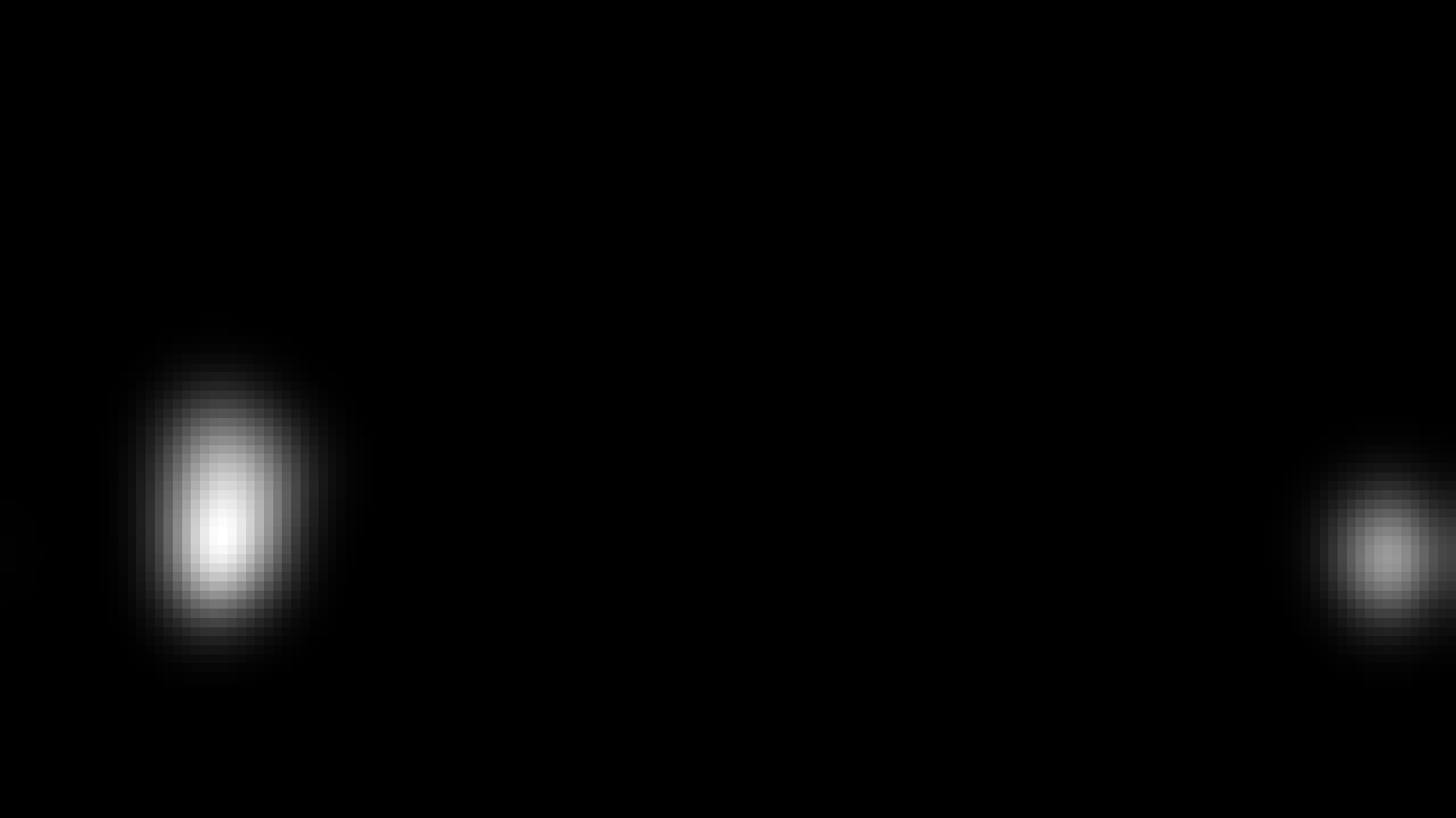}}
\hspace{0.01\columnwidth}
\subfloat{\includegraphics[width=0.23\columnwidth]{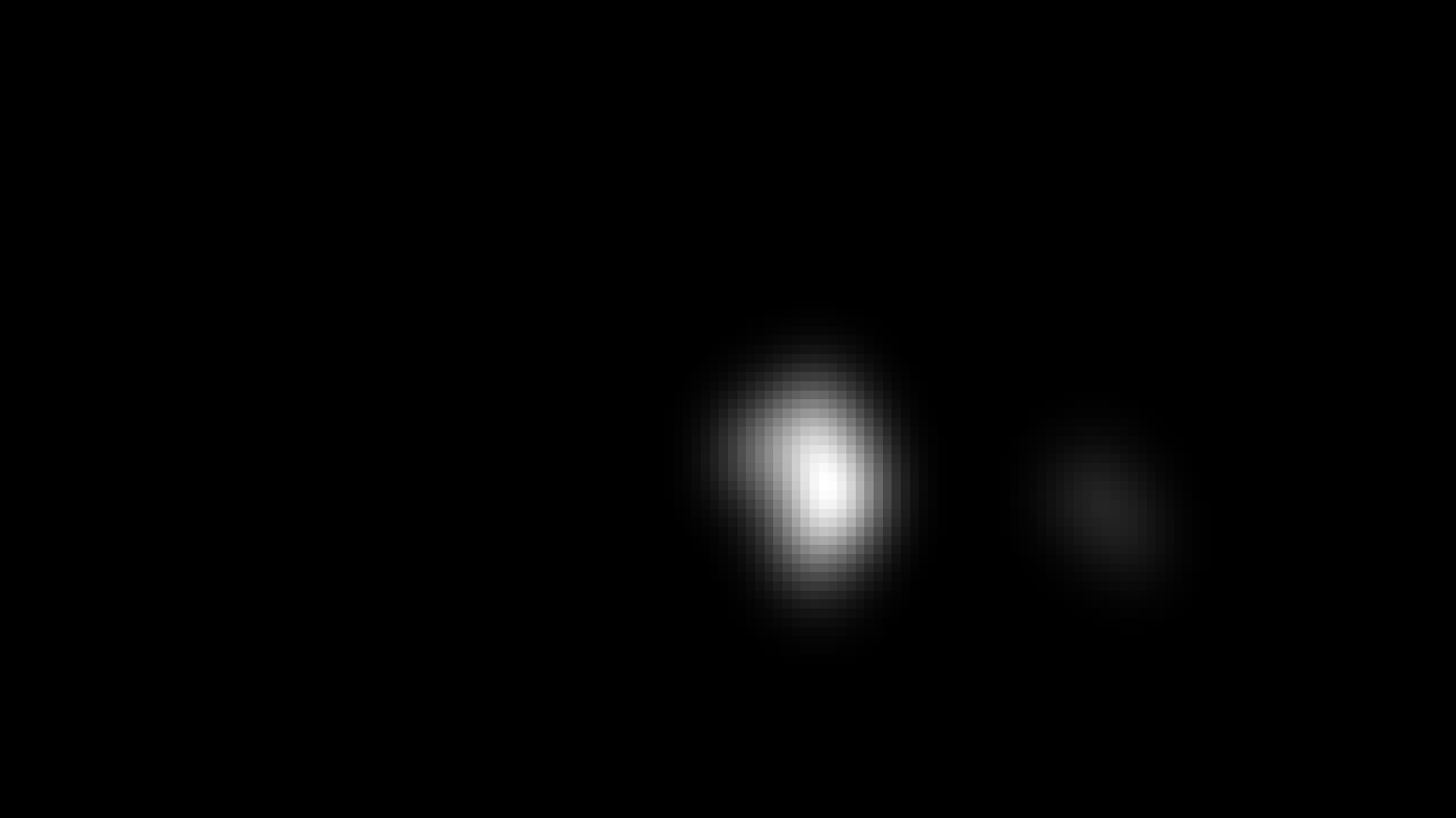}}
\hspace{0.01\columnwidth}
\subfloat{\includegraphics[width=0.23\columnwidth]{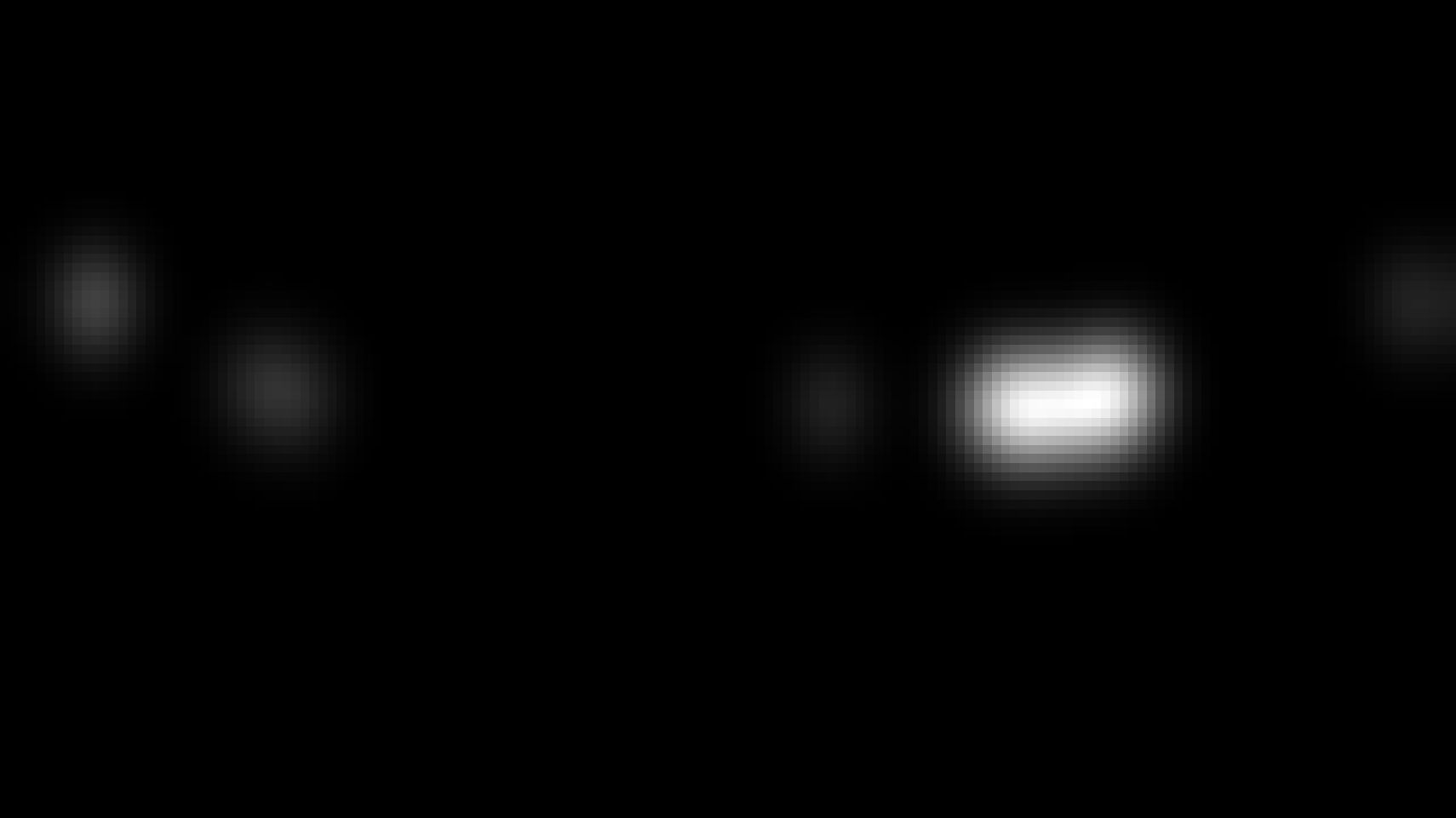}}\\
\caption{The saliency maps (bottom) uniquely identifies 360\degree\ videos (top) through regions of interest.
}
\label{fig_motivation_saldistinct}
\vspace{-2pt}
\end{figure}

\begin{figure}[t]
\centering
\subfloat{\includegraphics[width=0.45\columnwidth]{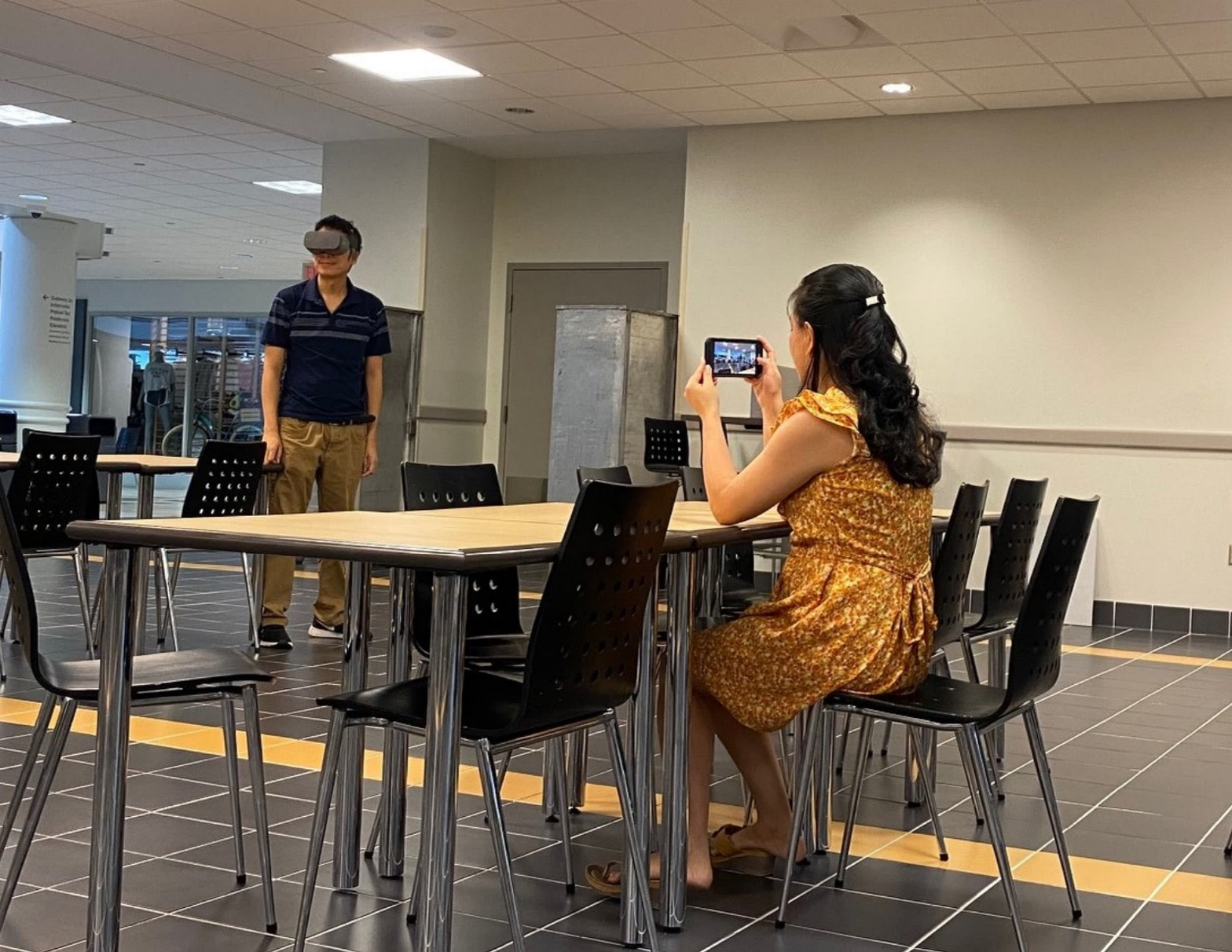}
\label{fig_illstration_attack1}}
\subfloat{\includegraphics[width=0.45\columnwidth]{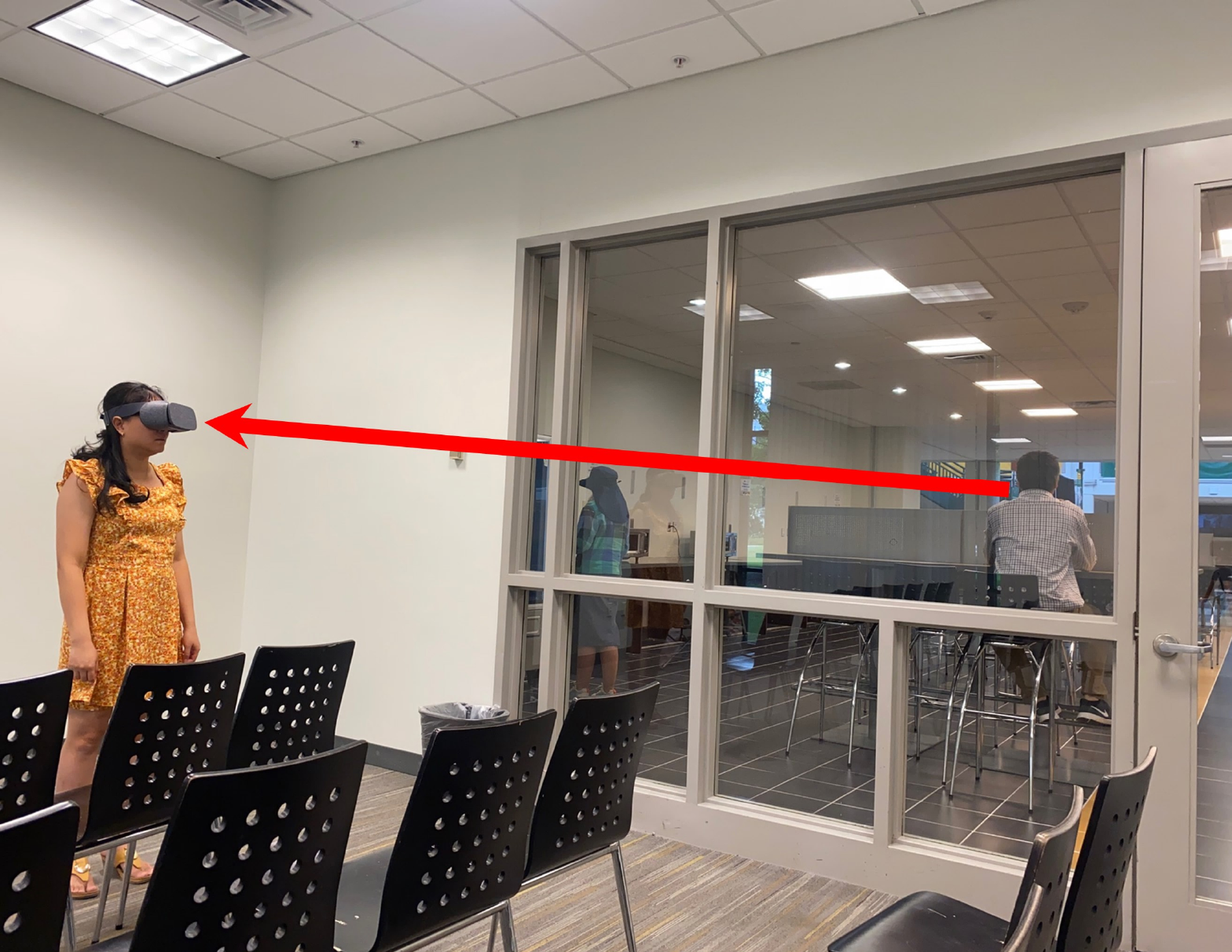}
\label{fig_illstration_attack2}}
\caption{Sample attack scenarios -- recording with a smartphone rear camera (left) and front camera (right).}
\label{fig_illustration}
\vspace{-1pt}
\end{figure}

The saliency of a video can be quantified by {\it saliency maps}. A saliency map is a 2D heat map that indicates the salient regions in a video frame. More salient regions where users stay for a long time are assigned higher saliency values and represented by brighter pixels in the saliency map, and vice versa for less salient regions that users glance through. The saliency map of an image can be generated by saliency detection, an established technology that has been widely used in surveillance \cite{yubing2011spatiotemporal}, advertising \cite{leveque2019eye}, and content generation \cite{wang2015saliency, jiang2021saliency}. As shown in \autoref{fig_motivation_saldistinct}, the saliency maps in the bottom row signify salient regions in the frames of four 360\degree\ videos in the top row. These saliency maps contain distinctive information about a visual scene that uniquely identifies the 360\degree\ videos. For instance, the man on the stage in the left-most image is the most conspicuous. Note that saliency maps remain effective in fingerprinting video frames even when they contain multiple salient regions. For example, the second-from-the-left saliency map in \autoref{fig_motivation_saldistinct} has two salient regions. It is possible for different users to attend to either the left or right region, with the likelihood being proportional to the brightness of the salient region. However, regardless of which region the users focus on, their head movement traces would still be correlated with this double-region saliency map rather than a saliency map of another video frame that has no overlap with their head movements. Essentially, the saliency map essentially captures the collection of regions with frequent head movements.

As for a video, the saliency map changes over time for each frame, following the progress of the content. The saliency fingerprint of a video, i.e., a sequence of saliency maps, becomes even more distinct. Therefore, we can conclude that it is \textit{feasible} to utilize saliency maps to fingerprint a 360\degree\ video.

\section{Overview of \sysname}
\label{sec:overview}
\label{overviewattack}

\subsection{Threat Model}
\label{sec:overview:threat}

We consider an adversary with a library of target 360\degree~videos whose viewers can be exploited for malicious purposes, e.g., pushing agendas or conducting blackmail. The attacker wants to know if a victim is viewing one of these videos and, if so, infer the specific video title. Existing studies~\cite{schuster2017beauty,reed2016leaky,gu2019traffic,maiti2019light,xu2014watching} have confirmed the sensitivity and privacy associated with such video title information. In our attack scenario, the victim is viewing a 360\degree\ video using an HMD
in a public space (e.g., a library\cite{statetechmagazine2019} or an airport\cite{airporttechnology2018}). Watching 360\degrees videos publicly with VR headsets has become popular due to the low purchasing cost \cite{urlyoutube360}. For example, 100\% libraries from the Association of Research Libraries provided VR headsets for in-library usage \cite{urlpurduelibrary}. There is also an increasing number of users watching 360\degrees videos during a flight \cite{urlmedium, urlredditvr}.
At the same time, the attacker is recording videos of the victim's head movement using a filming device, such as a smartphone or a digital camera (examples shown in~\autoref{fig_illustration}). The captured videos are referred to as \observations and used to infer the 360\degrees video viewed by the victim. This scenario is inspired by prior works where the attacker captures shoulder surfing recordings to infer the victim's unlock patterns on Android~\cite{ye2017cracking,khan2018evaluating,von2015easy}. However, our case is more plausible because the victim’s eyes are fully covered by the HMD and thus the victim is less vigilant about the proximity. The attacker may choose to place a hidden camera
to record the victim's head movement
to further reduce suspicion.

We assume the attacker has physical access to the location
where the victim is viewing 360\degree\ videos and can capture the recording. Also, we assume that the attacker knows the type of HMD the victim is using, which is trivial to obtain since the attacker can visually recognize it. 
The HMD type information will be used to tune the system components of \sysname. We do {\it not} assume any other capabilities for the attacker. For example,
the attacker cannot lure the victim into downloading and installing a malicious App \cite{slocumgoing}, 
and the attacker cannot sniff the network traffic 
coming from the victim's device. 

\subsection{Challenges and Key Components}
\label{sec:overview:challenges}
To identify 360\degree\ videos from recordings of victim head movement, we need to tackle the following  challenges.
\begin{packeditemize}
    \item \bheading{Extracting Head Movement Traces in VR}. 
    Head movement tracking in VR presents a distinct challenge compared to traditional head pose estimation in 2D scenarios~\cite{murphy2008head}. VR head movement encompasses omnidirectional viewing directions and the user's face is covered by the HMD. As a result, prior head pose estimators \cite{ruiz2018fine,kellnhofer2019gaze360,bermejo2020eyeshopper} assuming that head poses and facial features are explicitly displayed in front of the camera in a limited range of orientations would be ineffective.
    \item \bheading{Identifying Videos by Trace-Fingerprint Matching}. 
    Another challenge is that the video title is hidden inside the spatial-temporal correlation between two different data sources -- the head movement trace and the video fingerprint. Unlike previous video identification attacks that utilize a univariate time series, e.g., network traffic \cite{gu2019traffic} or light effusions \cite{xu2014watching}, to fingerprint a video and match it to the victim's corresponding time series, fingerprinting a video by fluctuating head movement traces is ineffective for trace-fingerprint matching and video identification.
\end{packeditemize}

\begin{figure}[!t]
\centering
\vspace{-1pt}
\includegraphics[width=0.48\textwidth]{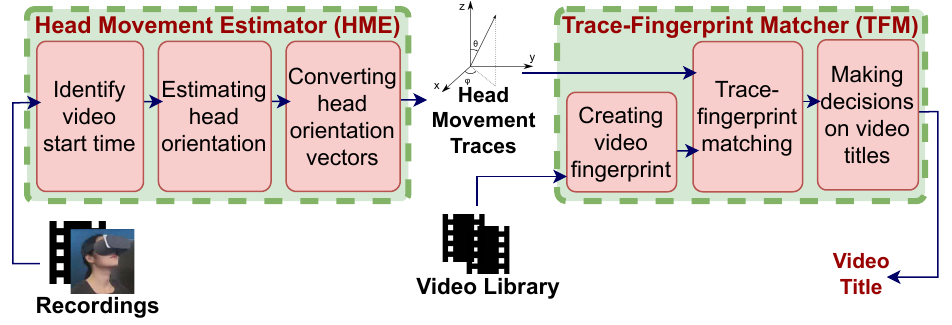}
\vspace{-15pt}
\caption{Overview of the proposed \sysname system.}
\label{fig_overviewAttack}
\vspace{-4pt}
\end{figure}

\sysname overcomes these challenges through its two components as shown in \autoref{fig_overviewAttack}. First, the head movement recording of the victim while viewing an unknown video is passed to the HMD-based head movement estimation scheme (HME). This scheme accurately identifies and extracts visual head movements during the omnidirectional viewing. The extracted movements are then converted into a head movement trace, a time series of coordinates tracking the head orientations relative to the VR coordinate system throughout the viewing session. Second, the head movement traces are fed to the video saliency based trace-fingerprint matching framework (TFM). The \VideoIdentifier matches the victim's trace with each video fingerprint in the library of target videos and produces a video identification decision. 

\section{HMD-based Head Movement Estimation (HME)}\label{section_extract_virtual_head_movement}
\label{sec:hme}
This section presents \sysname's head movement estimation scheme. \sysname identifies the start of video playback from the recording, extracts head orientations using the proposed DCNN, and generates the head movement trace after representing the head orientations in the VR coordinate system.

\subsection{Identifying the Start of Video Playback}\label{sec:hme:start}
A prerequisite to launching \sysname is determining if the victim is viewing a 360\degree\ video or engaging in other VR activities. Filtering the recording to obtain the 360\degree\ video viewing session is necessary because the recording may contain redundant information generated when the victim is using other VR Apps. If the victim is viewing a 360\degree\ video, it is also critical to identify the moment when the 360\degrees video starts playing on the HMD screen. After this moment, the victims will only move their heads based on the video content. Hence, this moment should be identified as the start of the head movement trace for launching the attack.

This challenge can be overcome by directly inspecting the recording due to the unique interaction pattern in 360\degree\ video viewing sessions. Initiating the playback requires the user to navigate to the desired video and ``click'' it for playback. Despite the differences in user interfaces in different HMDs, the navigation and selection of the video are always executed by a set of head and hand actions, ending with a press on the hand controller for starting the playback. After this final clicking, the viewing session starts and the victim only moves his/her head to explore the 360\degree\ video without further operation on the controller. This head-only, hand-free interaction design ensures immersive user experience \cite{voigt2020influence, mcgloin2013video}, differing from other VR Apps that always require both hand controller and head movement. By recognizing this unique sequence of user actions through visuals and sounds in the camera recording, we can infer that the victim is viewing a 360\degree~video and pinpoint the moment of the final controller clicking as the playback start time. \sysname extracts all frames from this starting moment for a recording length of $T$ seconds. 

We then locate and crop the regions that are relevant to the attack in the filtered recording. This can be easily performed by off-the-shelf video editing software because victims interact with 360\degrees videos only through head movement and their body positions remain stable during the viewing. Consequently, the recording becomes a smaller version only containing the victim's head, shoulders, and HMD. This preprocessing reduces the input size and thus the computation cost. It also eliminates noise in the background pixels, boosting the head movement estimation performance. After cropping, the recording is split into individual video frames before being sent to the head orientation estimation model.

\begin{figure}[!t]
  \centering
  \includegraphics[scale=.3]{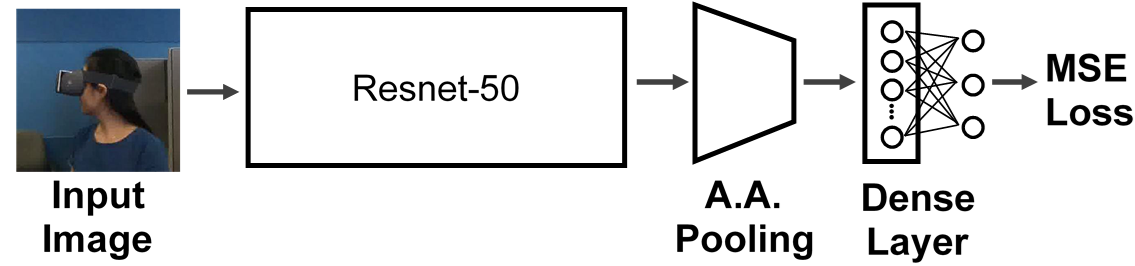}
  \vspace{-2mm}
  \caption{The \DNNEstimator.}  
  \label{fig_headposeEstimator}
  \vspace{-1pt}
\end{figure}
\subsection{Estimating Head Orientation}\label{sec_headpose_estimator}
This section details our head orientation estimation model's design, which analyzes the processed frames in the recording. The core of the head movement estimation (HME) component, the \DNNEstimator, transforms visual data from the \observation into numerical head orientation data. Previous estimation methods \cite{bermejo2020eyeshopper,ruiz2018fine,patacchiola2017head,zhu2016face} extracted head orientations from general non-HMD head pose images. By assuming that users mostly look straight ahead into the camera, they relied on facial landmarks to derive head orientations. However, these models fail when users' faces are obscured by HMDs and users turn their heads away from the camera's filming direction during 360\degrees video viewing. In \sysname, our \DNNEstimator is specifically designed to estimate HMD users' head orientations without the limited range of head movements assumed by prior works. It focuses on learning the visual features of HMDs, which are more discernible than facial landmarks, particularly when users deviate 45\degrees or more from the camera's filming direction. Therefore, our model can support the estimation of the full spectrum of head movements in 360\degrees video viewing. 

\bheading{Estimation Model Design.}
The \DNNEstimator architecture, depicted in \autoref{fig_headposeEstimator}, is a deep convolutional neural network (DCNN) comprising a base network and a decision layer.
The base network explores pixel values of the input frame. It extracts and learns the representation of meaningful features for the head orientation estimation. The decision layer estimates the head orientation based on the output of the base network.
Inspired by the ResNet-50 model \cite{he2016deep}, our base network consists of 48 stacked convolutional layers and an adaptive average pooling layer. The ability to learn complicated visual patterns of HMDs through this highly non-linear structure is our key advantage in head orientation estimation compared with previous approaches using hand-crafted features. The adaptive average pooling layer receives a set of feature maps from the base network. It then calculates the mean values of each feature map and returns a flattened vector.
By fixing the length of the flattened vector, our design can handle different sizes of input frames. 
Finally, the dense decision layer maps the flattened vector to a \textit{head orientation vector}, a 3D vector representing the head orientation in 3D spaces. 
\begin{figure}[!t]
\centering
\begin{tabular}{ccc}
\subfloat{\hspace{0cm}\includegraphics[width=0.13\textwidth]{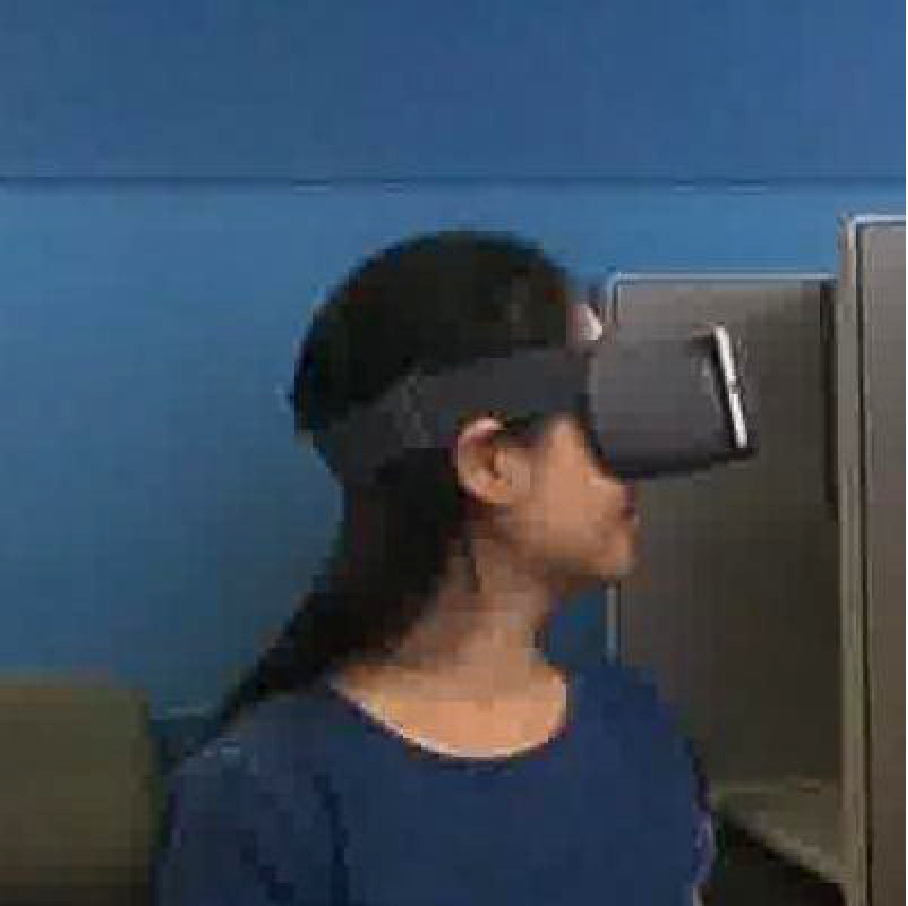}} & 
\subfloat{\hspace{0cm}\includegraphics[width=0.13\textwidth]{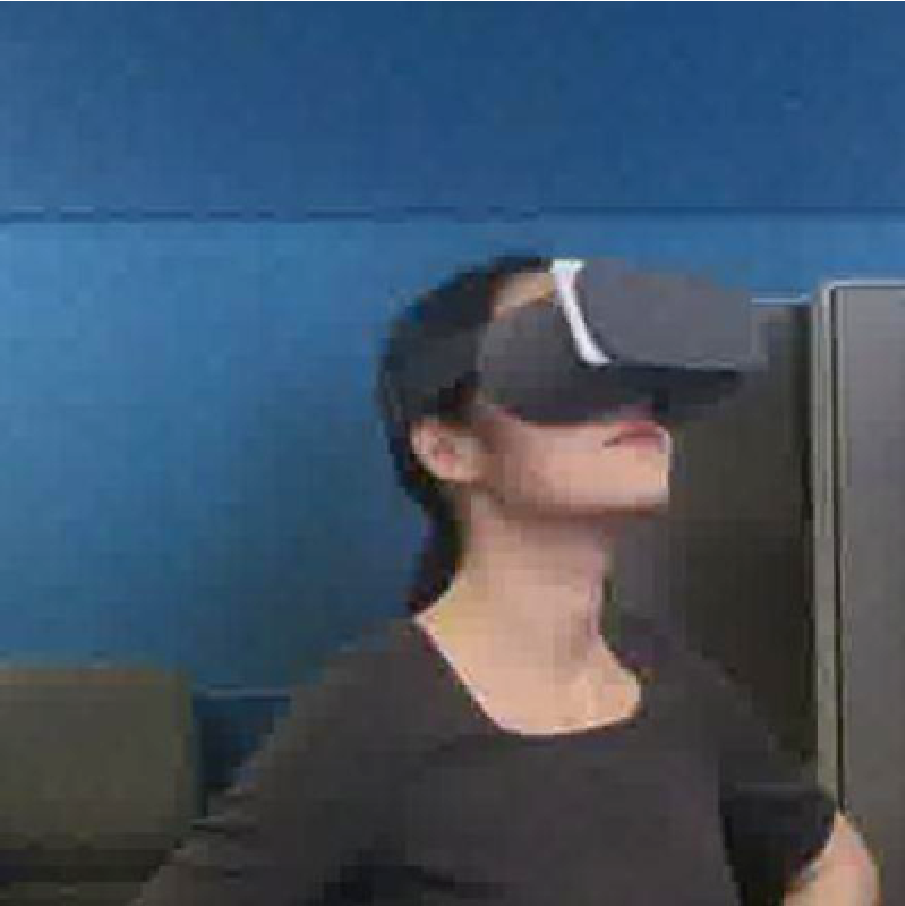}} &
\subfloat{\hspace{0cm}\includegraphics[width=0.13\textwidth]{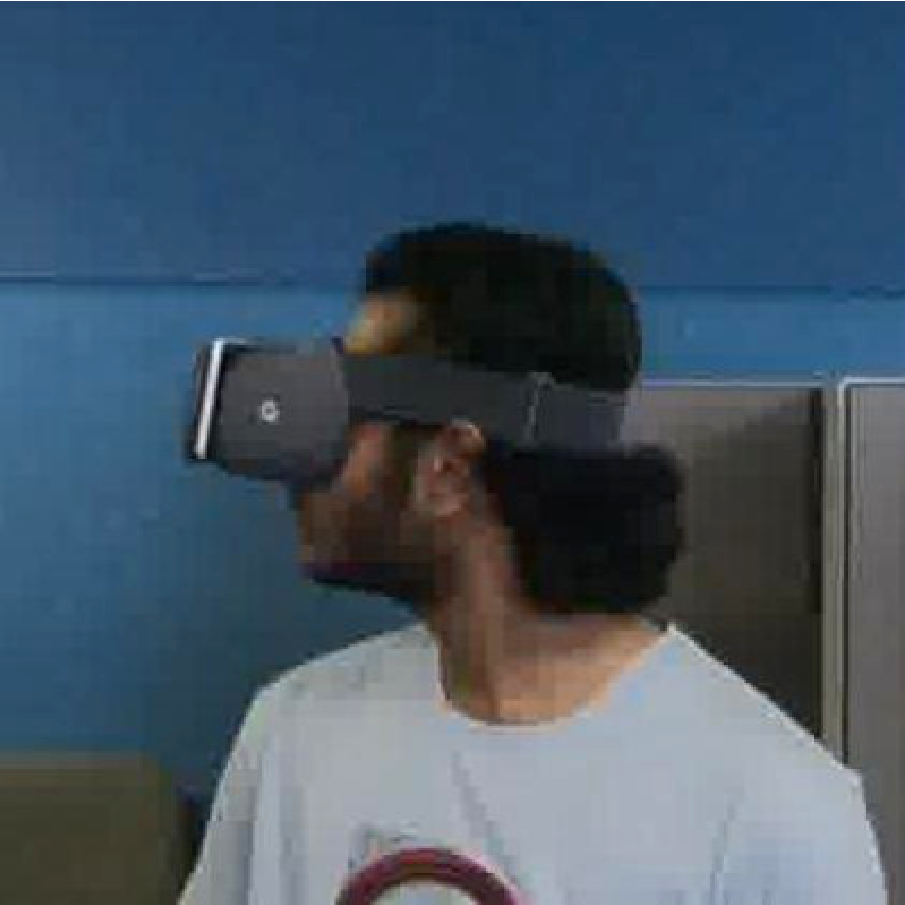}}\\[-3ex]
\subfloat{\hspace{0cm}\includegraphics[width=0.13\textwidth]{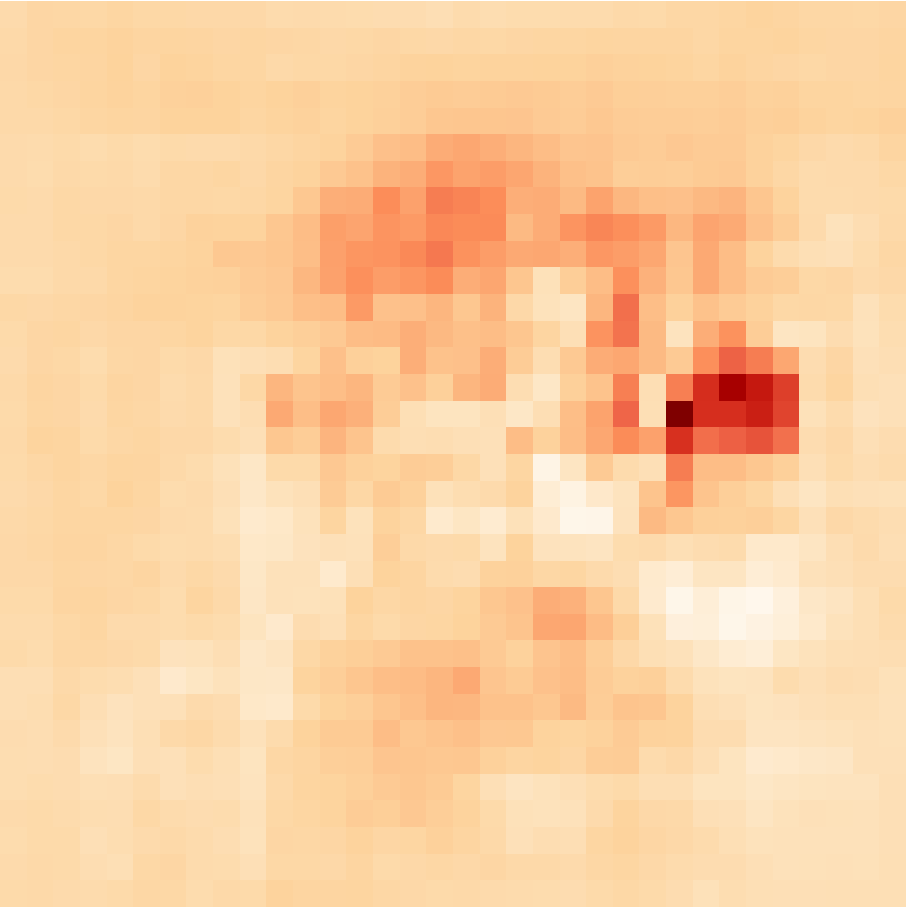}} & 
\subfloat{\hspace{0cm}\includegraphics[width=0.13\textwidth]{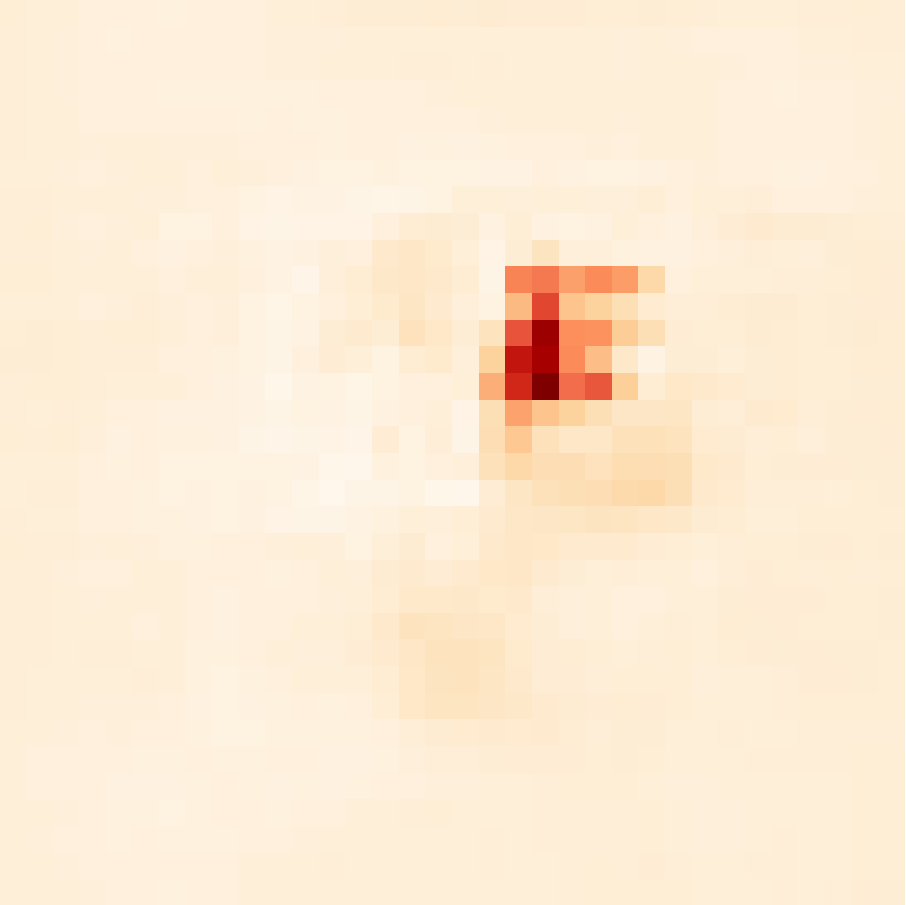}} &
\subfloat{\hspace{0cm}\includegraphics[width=0.13\textwidth]{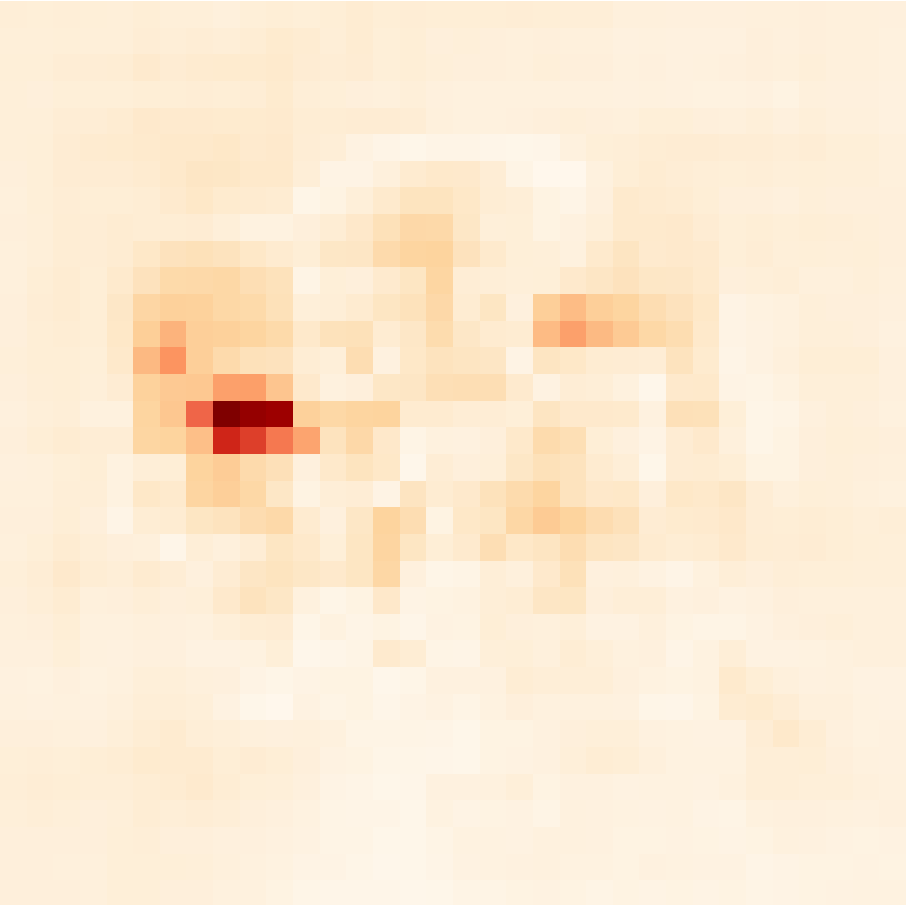}}\\
\end{tabular}
\caption{The head orientation estimation model focuses on learning HMD features (dark red areas in the bottom row) and can generalize well to victims not presented in the training.}
\label{fig_exp_hpestimator_illustrate}
\vspace{-1pt}
\end{figure}

\bheading{Model Training and Operation.}
We use Mean Squared Error (MSE) loss to guide the learning of the \DNNEstimator. It calculates the mean squared distances between the estimated and the ground truth head orientation vectors and forces the model to push its estimation to the ground truth. Formally, MSE is defined as:

\vspace{-10pt}
\begin{equation}
 \label{eq_hme_loss}
 \mathcal{L} = \frac{1}{3}((x - \hat{x})^2 + (y - \hat{y})^2 + (z - \hat{z})^2) 
\end{equation}
\vspace{-10pt}

where $v = (x, y, z)$ is the estimated head orientation vector extracted from input frame $f$, and $\hat{v} = (\hat{x}, \hat{y}, \hat{z})$ is the ground truth head orientation associated with $f$.

Since the above head orientation estimation model is not a personalized model designed only for
a particular victim, attackers can obtain training data offline by rotating their heads to different angles with respect to the camera's filming direction, just as normal users. They can capture the camera recording of their own head movements by using the same type of HMD as the victim. Meanwhile, the ground truth head orientations can be collected by the HMD sensors. Once the model is trained, it can then be used to estimate the head orientation of the victim who is not in the training set, i.e., a testing user.  \autoref{fig_exp_hpestimator_illustrate} illustrates that the trained model focuses on learning the features of HMDs and can be well generalized to new testing users. Through the occlusion method \cite{zeiler2014visualizing}, the heat maps at the bottom row highlight the regions affecting the estimation performance for the testing frames in the top row. These regions contain HMDs (darker red) rather than heads, ears, or jaws (lighter red) and are consistent across victims wearing similar HMDs. After applying the model to each video frame in the recording, a sequence of head orientation vectors is eventually generated.

\subsection{Converting Head Orientation Vectors}\label{sec_convert_headoren}

The proposed estimation model generates a sequence of head orientation vectors, but these cannot directly form a head movement trace because they are referenced to the recording camera's filming direction, i.e., the \textit{camera-based coordinate system}. This differs from the \textit{VR coordinate system} used in VR Operating Systems (OS) for tracking head movements and rendering VR content. As a result, an extracted vector may point to a totally different direction in the VR OS. Hence, \sysname must convert these vectors to the VR coordinate system before inferring a victim's regions of interest and video title.

\begin{figure}[!t]
  \centering
  \includegraphics[width=.32\columnwidth]{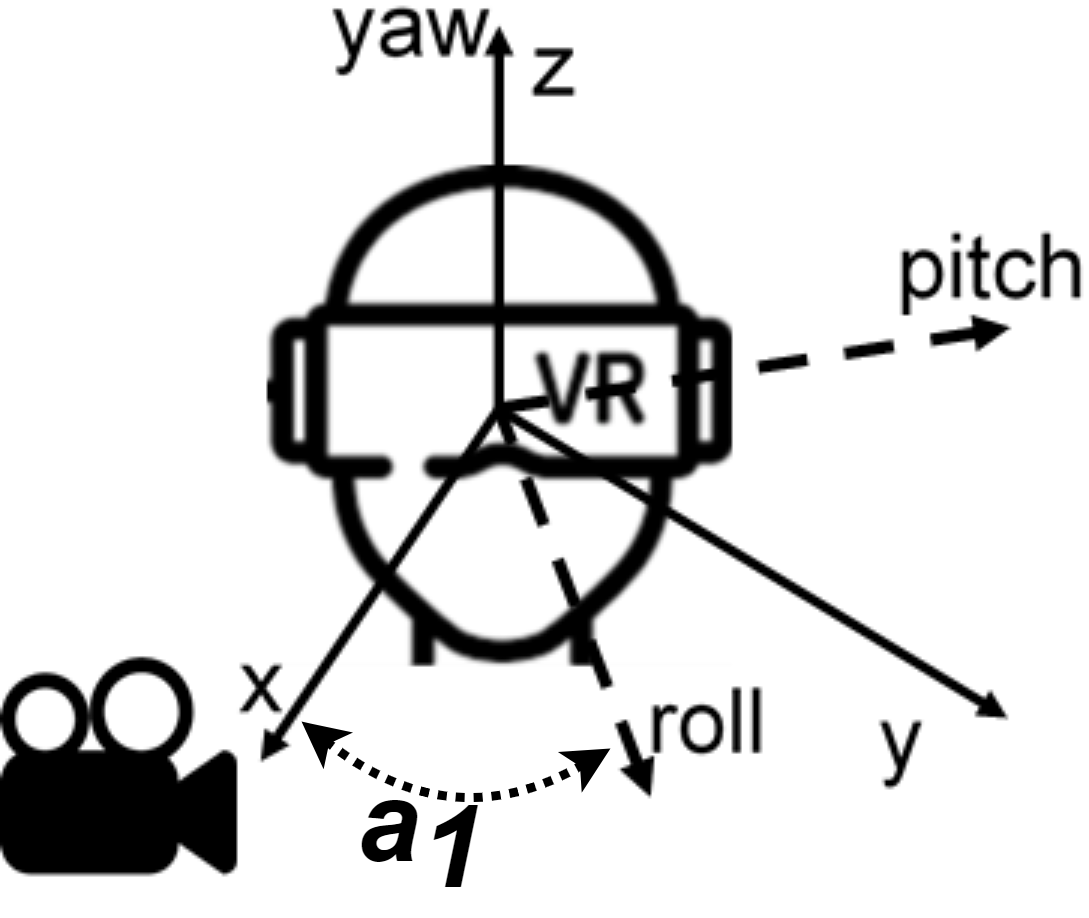}
  \caption{Difference between the camera-based coordinate system (solid lines) and the VR coordinate system (dash lines).} 
  \label{fig_hmdcoord}
  \vspace{-1pt}
\end{figure}

\autoref{fig_hmdcoord} illustrates the difference between the two coordinate systems. It can be seen that both systems have an axis pointing straight up (the $z$ and yaw axes), but their reference axes are distinct. The reference axis in the camera-based coordinate system (the $x$ axis) aligns with the camera's filming direction and can be visually identified in the camera recording. On the other hand, the reference axis of the VR coordinate system (the roll axis) is represented in the VR virtual world and cannot be observed in the physical world by the attackers.

To locate the VR coordinate system, we first identify the roll axis by leveraging the fact that the VR coordinate system is not fixed in the physical world but is always reset by the VR OS at the beginning of a 360\degree\ video viewing session. Upon reset, the roll axis is configured to overlap with the projection of the head orientation vector onto the $x$-$y$ plane. By extracting the head orientation vector at the reset, i.e., at the starting moment of the video playback, we locate the roll axis in reference to the camera-based coordinate system.

\sysname can then convert a head orientation vector for it to be represented in the VR coordinate system. As shown in \autoref{fig_hmdcoord}, the two systems already have two axes aligned, namely the yaw and $z$ axes. Thus, the conversion becomes the angular rotation along the $z$ axis for the $x$ and roll axes to align. As discussed above, the offset angle between the $x$ and roll axes can be derived when locating the roll axis. Without loss of generality, we denote the first frame at the beginning of the viewing session as $f_1$ and the head orientation vector in the camera-based coordinate system extracted from $f_1$ as $v_1$, $(x_1, y_1, z_1)$ being the coordinate of $v_1$. Let $P$ be the plane formed by the intersection of the $x$ and $y$ axes. Then the offset angle $a_1$ between the $x$ axis and roll axis (the projection of $v_1$ onto $P$) can be expressed as $a_1 = arctan(\frac{y_1}{x_1})$.
Given the offset angle $a_1$, the mapping $q_1$ is the quaternion rotation operation \cite{ben2014tutorial} rotating a 3D vector for $-a_1\degree$ along the yaw axis. Applying $q_1$ to a head orientation vector $v$ in the camera-based coordinates produces the corresponding representation in the VR coordinate system $u$.

\section{Video Saliency-based Trace-Fingerprint Matching (TFM)}\label{sec:identifier}
This section details \sysname's approach to inferring video titles through trace-fingerprint matching. \sysname generates video saliency-based fingerprints for 360\degree\ videos targeted by the attacker, and matches the victim's head movement trace obtained from the HME component with each fingerprint to infer the video being viewed.

\subsection{Creating Fingerprints for 360\degree\ Videos}
\label{sec:identifier:create}
An essential offline task for attackers before making the video identification decision is to build a library of videos of interest and create fingerprints for these videos. These videos of interest can be carefully selected based on the malicious purpose. For example, in the case of blackmailing, sensitive videos that the victim is not supposed to view can be put in the library.  
As discussed in \autoref{sec_mot_how}, a head movement trace of a single user is insufficient to fingerprint a 360\degree\ video because the head movement time series generally fluctuates. However, users tend to agree on the regions of interest in 360\degree\ videos. This fact drives the similar head movement patterns across users that center around salient regions of a video. Therefore, we propose to utilize video saliency maps to fingerprint a 360\degrees video. \looseness=-1

A saliency map can be derived offline by collecting and processing head movement traces of a group of users, but it would be too expensive to collect subjective human data for every 360\degrees video targeted by the attackers. \sysname instead uses an established saliency detection model to generate the saliency map. Thanks to the recent development of deep learning, salient regions of human users on 2D videos and 360\degree\ videos can be reliably and accurately detected \cite{nguyen2018your,jiang2015salicon}. These state-of-the-art saliency detectors have been trained on a large amount of data and can identify the salient regions of new unseen videos. To fingerprint a video, \sysname performs saliency detection on all its frames through a state-of-the-art model. The output of the model, a sequence of saliency maps, is a spatial-temporal video signature reflecting the progress of human attention through time and is stored as the video fingerprint. This procedure is repeated for all library videos.

\subsection{Matching Victim Trace with Fingerprints}
\label{sec:identifier:tfm}
Unlike the majority of previous works that only processed time series to identify video titles \cite{schuster2017beauty, ling2019know, maiti2019light, xu2014watching}, \sysname needs to handle input of different modalities to infer the video title. The victim's head movement trace is a time series of head orientation vectors, whereas the fingerprint of a 360\degree\ video is a sequence of 2D saliency maps. \sysname's \VideoIdentifier must establish the spatial correlation between the victim's head movement and the video's salient regions, as well as comprehend the head movement patterns in relation to the progression of the video's salient regions.
Such tasks are challenging for traditional machine learning models hand-crafted by domain knowledge \cite{schuster2017beauty,reed2016leaky,gu2019traffic}. To overcome this, we propose a DCNN to capture the interactive patterns between the two input modalities, effectively matching the victim's head movement trace to a library of video fingerprints for video identification.

\bheading{Input Preprocessing.}
Since the victim's head movement trace and the video fingerprint are two different modalities, they cannot be analyzed simultaneously. We propose to first transform the head movement trace into a sequence of 2D head orientation maps. Let $\mathcal{V}$ be the head movement trace and $\mathcal{M}$ be the video fingerprint, where $\mathcal{V}={v_1, .., v_N}$ is a list of 3D head orientation vectors in the VR coordinate system and $\mathcal{M} = {m_1, ..., m_N}$ is a list of saliency maps. Each saliency map $m_i$ is a 2D heat map of size $W \times H$. 
We create the head orientation maps through equirectangular projection \cite{ray2018low}, a prevalent approach for projecting 3D spherical data onto a 2D plane. We adopt this method because modern 360\degree\ video saliency maps are marked on equirectangular 360 frames \cite{nguyen2018your,nguyen2019saliency}. This projection aligns both input modalities. For a 3D head orientation vector $v_i$, we find the azimuth and altitude angles, $\theta_i$ and $\phi_i$. Then, the projected point $v_i'=(w_i,h_i)$ on the equirectangular of size $W \times H$ can be derived using the following formulas \cite{nguyen2019saliency},

\vspace{-8pt}
\begin{equation}
 \label{eq_equrec}
w_i = (\frac{\theta_i}{360})W, \ \  
h_i = (\frac{1 - sin(\phi_i)}{2})H. 
\end{equation}
\vspace{-10pt}

After this conversion, we obtain a sequence of head orientation maps of size $W \times H$ where the head orientation is indicated by $\mathcal{{V'}} = ({v'}_1, ..., {v'}_N)$.

We then combine the transformed head orientation maps with the video fingerprint into a unified input. Given that both head orientation maps and saliency maps are associated with timestamps that signal their temporal position in the video, we combine the corresponding maps of the same timestamp. 
The result is a sequence of pairs of head orientation maps and saliency maps that can be processed by our DCNN. The corresponding regions in both head orientation maps and saliency maps are aligned. Consequently, the DCNN better learns the spatial relevance between the victim's head orientation and the video's salient regions.

Due to the computational intensity of the DCNN, processing the entire sequence of head orientation and saliency maps can be inefficient. Yet, feeding a sparse number of maps to the DCNN may fail to present a continuous context of how the victim's head movement interacts with the video saliency. To balance efficiency and performance, we introduce $\tau$, a system parameter defining the time interval between two samples in the sequence. By adjusting $\tau$, \sysname can flexibly control the sampling interval of input data for the trace-fingerprint matching model. For instance, setting $\tau=1$ directs our system to sample the sequence of head orientation maps and saliency maps every second.

\begin{figure}[t]
 \centering
  \includegraphics[width=1.0\linewidth]{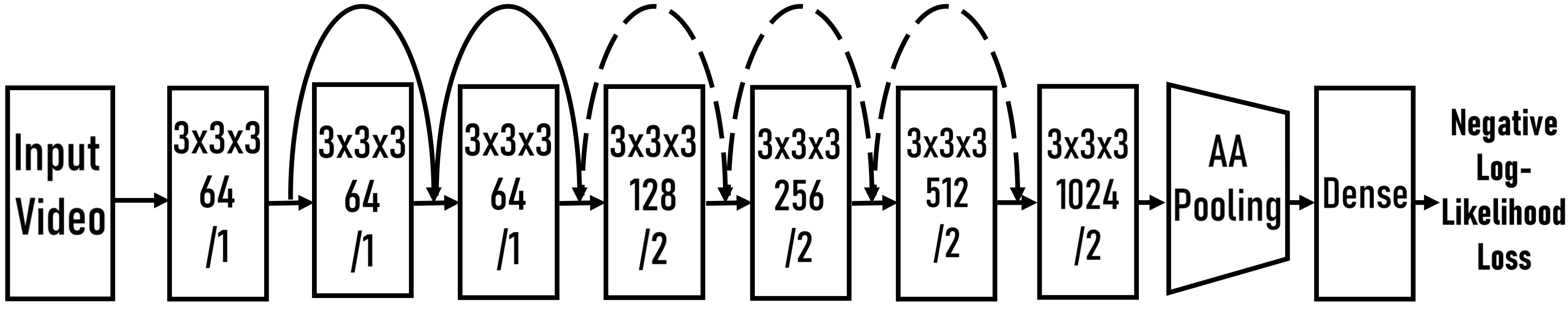}
  \caption{The DCNN architecture of the video saliency based trace-fingerprint matching model.}  
  \label{fig_videoIdentifier}
  \vspace{-1pt}
\end{figure}

\bheading{Matching Model Design.}
The matching between the preprocessed head movement trace (head orientation maps) and video fingerprints (saliency maps) can be regarded as a mapping problem, where the two input channels are mapped to a confidence score that indicates the correlation between them. 
The core of the trace-fingerprint matching model is a DCNN, as depicted in \autoref{fig_videoIdentifier}, composed of seven stacked 3D convolutional layers. Unlike traditional 2D convolutional layers, the 3D kernel convolves over several input maps at a time to learn both the localized spatial information in the head orientation maps and saliency maps, and the short-term temporal relationship between these maps. Stacking several 3D layers expands the neuron receptive fields, enabling the model to learn long-term information throughout the video. This design is consistent with the spatial-temporal structure of the two input modalities to match.

For each convolutional layer, there are three associated parameters, the kernel size, the number of kernels, and the stride.
The kernel size of $3 \times 3 \times 3$ is selected for all convolutional layers to reduce the computational cost.
Kernels having the stride value of 2 (denoted by $/2$) reduce the size of the intermediate maps by half.
We use residual connections \cite{he2016deep}, shown as U-shape arrows in \autoref{fig_videoIdentifier}, to connect the output of the previous layers with the output of the next layers. The connections facilitate the backward flow of gradients from the output back to the input, mitigating the impact of the gradient vanishing. 
Similar to the head orientation estimation model, the end of the DCNN is the adaptive pooling layer \cite{he2015spatial} converting the filtered maps of arbitrary sizes into a fixed flattened vector. The decision network is a dense layer of 1024 neurons mapping the flattened vector to a single confidence score in the range of $(0, 1)$.

To measure the difference between the output confidence score of the model and the ground truth, we use the Negative Log-Likelihood loss,

\vspace{-10pt}
\begin{equation}\label{eq_id_loss}
 L(n, \hat{n}) = -\hat{n}log(n) 
\end{equation}
\vspace{-15pt}

where $n$ is the output of the proposed model and $\hat{n}$ is the ground truth. If the head movement trace and the video saliency maps for matching indeed come from the same video, $\hat{n}$ has the value of 1. Otherwise, $\hat{n}$ has the value of 0. Since $\hat{n} \in \{0, 1\}$, the value of $L(n, \hat{n})$ can be either 0 or $-log(n)$.

\subsection{Making Video Identification Decisions}\label{section_making_decision}
As the trace-fingerprint matching model is not victim- or HMD-dependent, attackers can collect offline data of head movement traces and video fingerprints to train the proposed DCNN. The head movement traces can be collected by HMD sensors when attackers view target 360\degree\ videos and/or by directly using public head movement traces of popular 360\degree\ videos targeted by the attackers. The video fingerprints, i.e., the saliency maps, can be created by applying a high-performance saliency detector on the respective 360\degree\ videos. Once the model is trained, it can be used for video identification. To perform the attack, the victim's head movement trace is paired with each video fingerprint in the library and processed through the \VideoIdentifier. This outputs confidence scores that indicate the match between each video and the victim's head movement. \sysname then produces the top-$k$ candidates by selecting the $k$ videos with the highest confidence scores.

\section{Experimental Methodology}
\label{section_setup}
\subsection{Settings}
\label{sec:exp:settings}

\bheading{Apparatus.} The \sysname prototype comprises an HMD device, a GoPro HERO6 camera, and an offline processing desktop running on Ubuntu 18.04 equipped with a GeForce GTX 1080 Ti GPU. 
The GoPro6, which records videos in 1080p with a 12MP lens, is inferior in recording capability compared to modern smartphones such as iPhone 14 that is equipped with a 48MP lens and can produce 4K videos. We tested two cordless HMDs commonly used in VR studies \cite{grinyer2022effects, chen2021gestonhmd, speicher2021designers}, Google DayDream and Google Cardboard. Both have 100\degrees field of view similar to most modern VR headsets. The camera was situated 1.5 meters in front of the user at head level, and recorded at 1080p resolution. We took recordings in a well-lit office ($\sim$1000 lux as per \cite{gsa2020}) at a sampling interval of $\tau=0.8$. We extracted 60 seconds from each recording for our experiments. Half of them were extracted from the first 60 seconds, while the other half were segmented from a random point of the recordings. This default setting is maintained unless otherwise stated when evaluating specific factors.\looseness=-1

\bheading{Subjects.} We recruited 31 subjects (15 males and 16 females, aged from 19 to 36) for the user studies. Subjects are university students and people recruited through public channels, among which 18 wear glasses, 5 have no experience in VR, and 10 have little experience. All subjects gave consent to our IRB-approved study, which allows us to record human responses for VR system evaluation. Subjects stood in an open space while viewing the provided content. All 31 subjects participated in model training (\autoref{sec_offline_training}) while 17 subjects were randomly chosen to play the `victim' role during the data collection phase (\autoref{sec:exp:online}). The decision to work with 17 victims is attributed to the significant time commitment of playing a victim, aligning with the participant count in similar studies \cite{slocumgoing, wu2023privacy}. For training the HME, recruits were asked to follow specific instructions to collect head orientations. When playing the role of victims, users were instructed to engage freely with the 360\degrees videos without any additional guidance. All participants were kept unaware of the study's purpose.

\bheading{Video Library.}
We collected a library composed of the top 635 popular 360\degrees videos with the most views on YouTube, covering a diverse array of genres such as documentaries, music, sports, animations, wildlife, vlogs, travel, and gaming. Each genre contains a minimum of 30 videos and each video receives at least 40,000 views. We obtain the fingerprints of these videos by applying a state-of-the-art 360\degree\ video saliency detector \cite{nguyen2019saliency}. 

\subsection{Offline Model Training}
\label{sec_offline_training}

\bheading{The HME.}
To train the head orientation estimation model in the HME (\autoref{sec_headpose_estimator}), we first collected a dataset of camera recordings that captures different head orientations of the 31 subjects. Subjects were asked to view a target virtual object displayed on the HMD screen at different angles with respect to the camera filming direction. The head orientations during this one-second viewing session ranged from -179\degrees to 179\degrees in yaw and from  -55\degrees to 55\degrees in pitch. Note that subjects were not instructed to freely explore 360 videos because that may not cover the diverse head orientations needed for HME training.

Each subject repeated the viewing session 15 times for one HMD device. We then annotated the head orientation for each frame of the recordings as the ground truth. A total of $\sim$200,000 annotated head orientation frames were collected to train the model. To mitigate model overfitting and increase the variety of patterns the model can learn, we applied data augmentation to the training data. We randomly picked a method from horizontal flipping, adding random gray boxes, modifying brightness and contrast, and zooming in and out at the center, and then iterated through all frames \cite{shorten2019survey}. The total time to train the head orientation estimation model was approximately 5 hours. Details on creating the dataset can be found in the Appendix.

\bheading{The TFM.}
\label{section_creating_video_identifier}
As discussed in \autoref{section_making_decision}, the training process of the matching model uses head movement traces when viewing videos and fingerprints of these videos. An inherent advantage of this design is that TFM training does not rely on camera recordings of head movements, making it independent of the training and testing performance of the HME. This enables attackers to leverage publicly available head movement traces for TFM training, which offers more accurate data than the HME output and reduces the data collection efforts. We implemented this design by utilizing three existing datasets of head movement traces for 360\degree~videos \cite{dscorbillon, dslo, dswu}. These traces were collected by built-in HMD sensors for 24 videos, each with head movement traces from at least 48 users in a free-viewing context. In addition, we obtained the fingerprints of these videos by applying the aforementioned saliency detector \cite{nguyen2019saliency}. By combining a head movement trace and a saliency fingerprint as one sample, we created a training set of 1,152 positive samples and 23,392 negative samples, where \textit{positive} means the trace matches with the video title and \textit{negative} indicates no match. 

To mitigate the overfitting, we applied shifting augmentation~\cite{shorten2019survey} on head orientation maps, where the pixels indicating head orientations were randomly shifted up, down, left, or right to simulate the head orientation estimation errors. We leveraged an upsampling technique to mitigate the impacts of imbalanced data. The total training time was about 16 hours.

\subsection{Online Attacks}
\label{sec:exp:online}

Once the models in the HME and the TFM are trained, we launch \sysname to execute attacks and evaluate the performance of video identification. We conducted {\it cross validation}, wherein each round one user played the role of a victim and his/her data was excluded from the training set to ensure no overlap. We simulated the scenario where the victim viewed some videos in the target library, while the attacker aimed to determine which videos were viewed. We randomly selected 22 videos from the 635-video library for the victim to view, resulting in 22 camera recordings of head movements. For each recording, \sysname inferred the top-$k$ video titles out of the 635 potential options. We repeated this process for 17 victims and report the average results.

\section{Evaluation Results}\label{section_eva}

In this section, we present the extensive evaluation results of \sysname when the online attacks are launched. We start with validating the head movement estimation (HME) (\autoref{sec:eval:hme}). Next, we focus on evaluating the video identification attacks, including analysis of system parameters (\autoref{sec:eval:freq}), comparison with baseline approaches (\autoref{sec:eval:baseline}), and robustness analysis under various environmental factors (\autoref{sec:eval:robustness}). Finally, we evaluate the open-world identification (\autoref{sec:eval:open}).

\begin{figure}[tb]
\centering
\begin{subfigure}[b]{0.495\columnwidth}
\includegraphics[width=\textwidth]{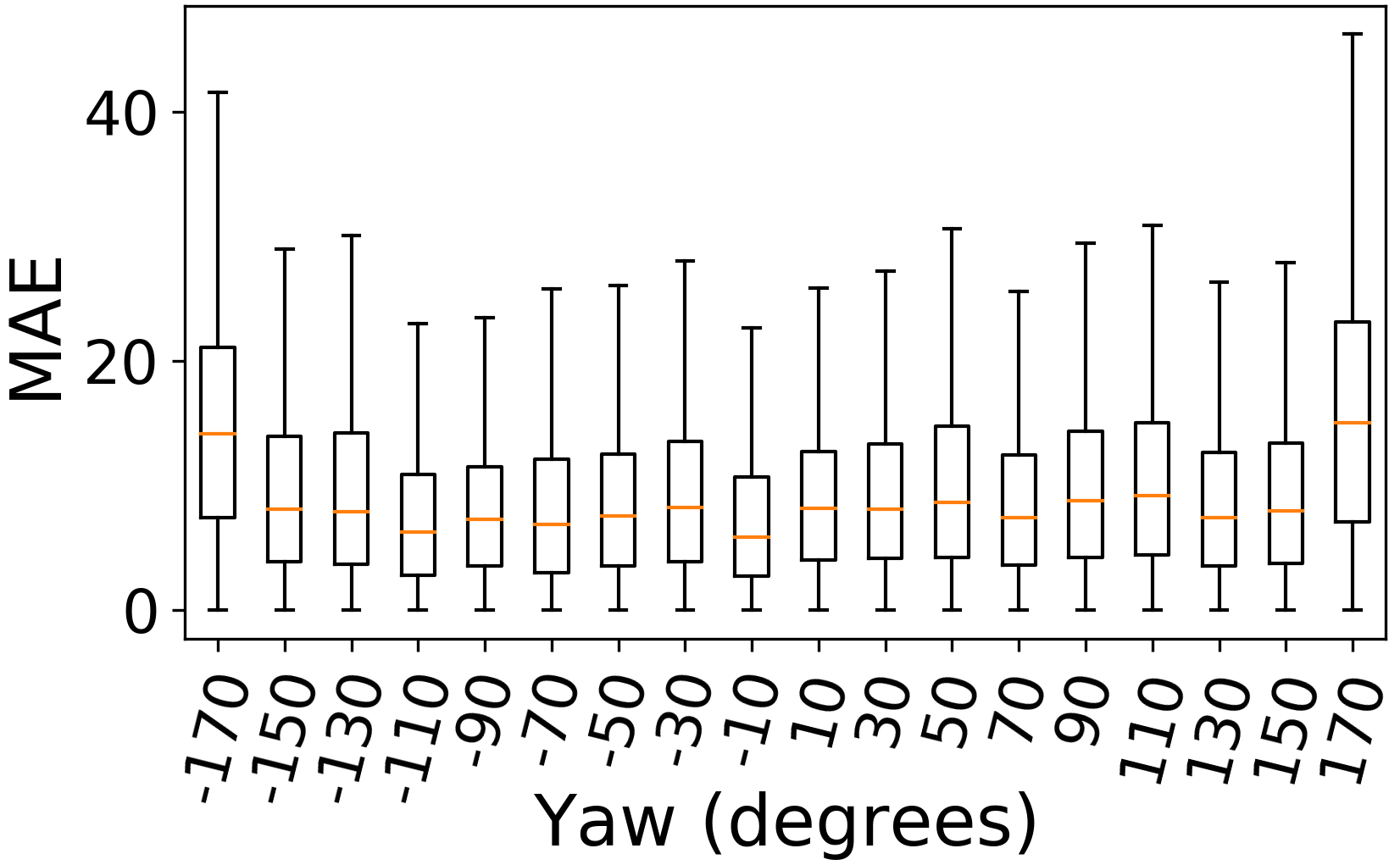}
\caption{Yaw}
\label{fig_exp_hpestimator_yaw_accuracy}
\end{subfigure}
\hfill
\begin{subfigure}[b]{0.495\columnwidth}
\includegraphics[width=\textwidth]{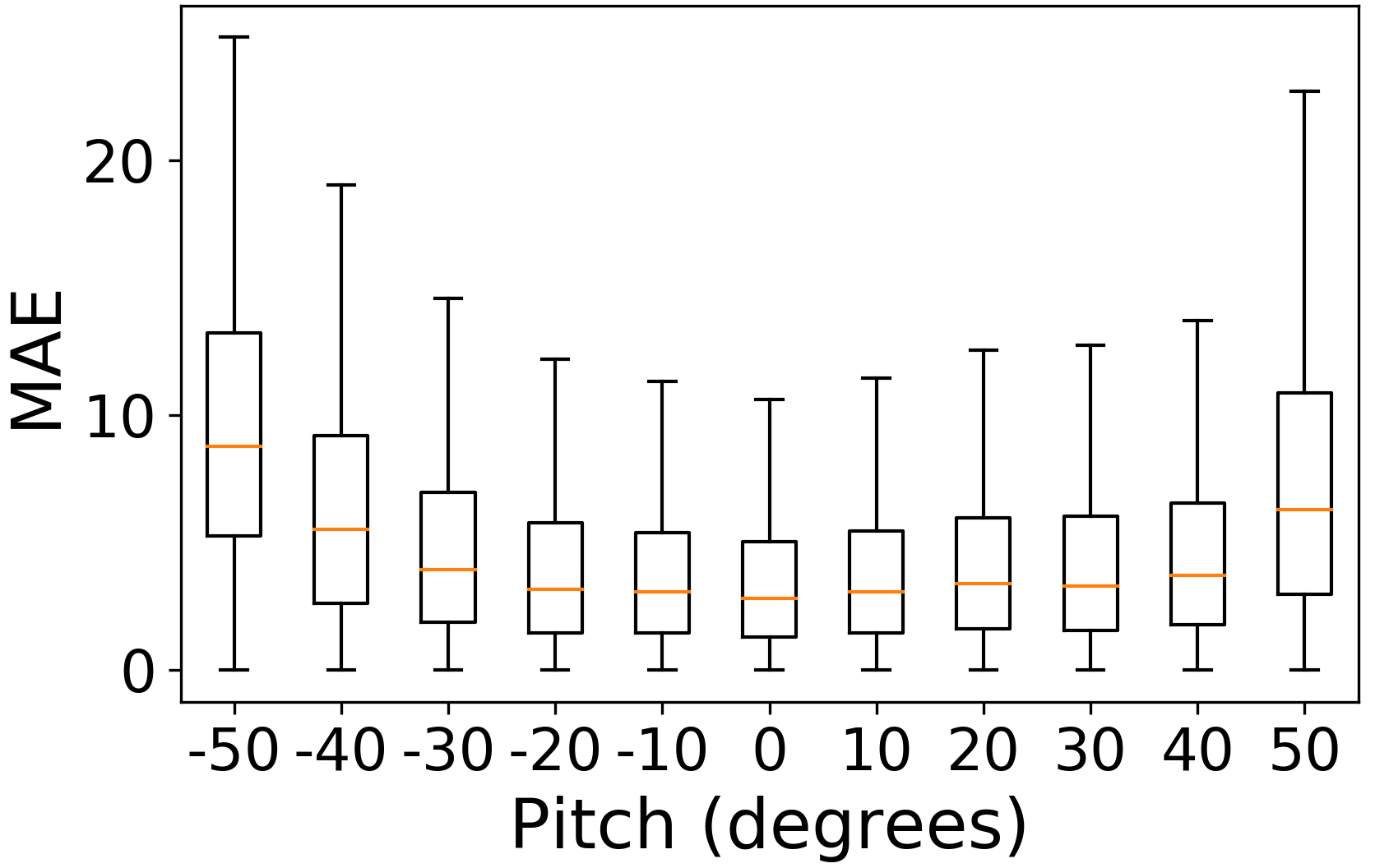}
\caption{Pitch}
\label{fig_exp_hpestimator_pitch_accuracy}
\end{subfigure}
\caption{Estimation errors of the HME in yaw and pitch.}
\label{fig_exp_hpestimator_accuracy}
\vspace{-4pt}
\end{figure}

\subsection{Validation of the HME} 
\label{sec:eval:hme}

\begin{figure*}[t]
  \centering
\begin{minipage}[t]{0.287\textwidth}\vspace{-9pt}
\includegraphics[width=.85\columnwidth]{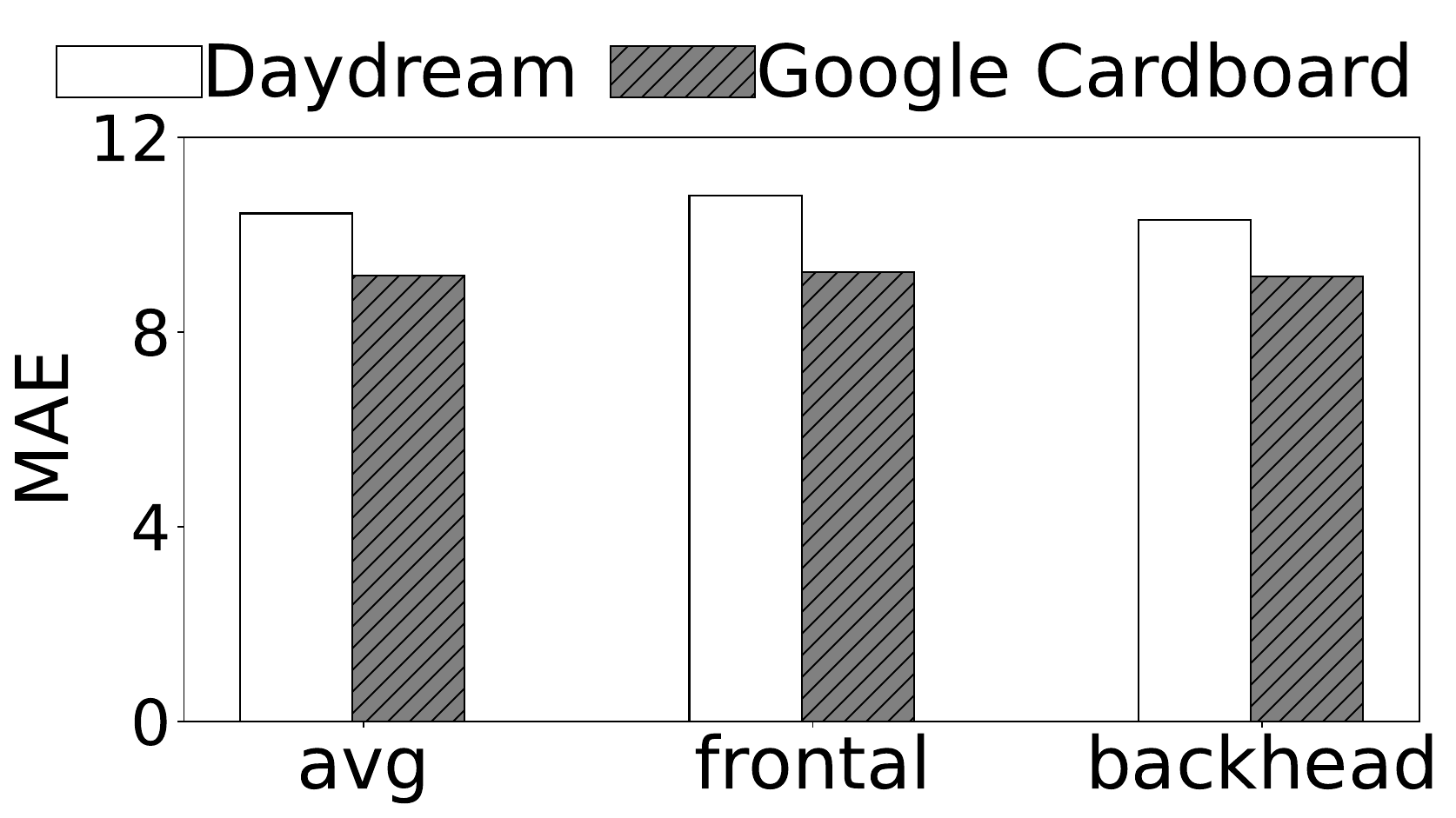}
\vspace{-5pt}
\caption{Estimation errors of the HME for different VR headsets.}
\vspace{-8pt}
\label{fig_exp_headsets}
\end{minipage}
\hspace{.5cm}
\begin{minipage}[t]{0.287\textwidth}\vspace{-0pt}
\centering
\includegraphics[width=.85\columnwidth]{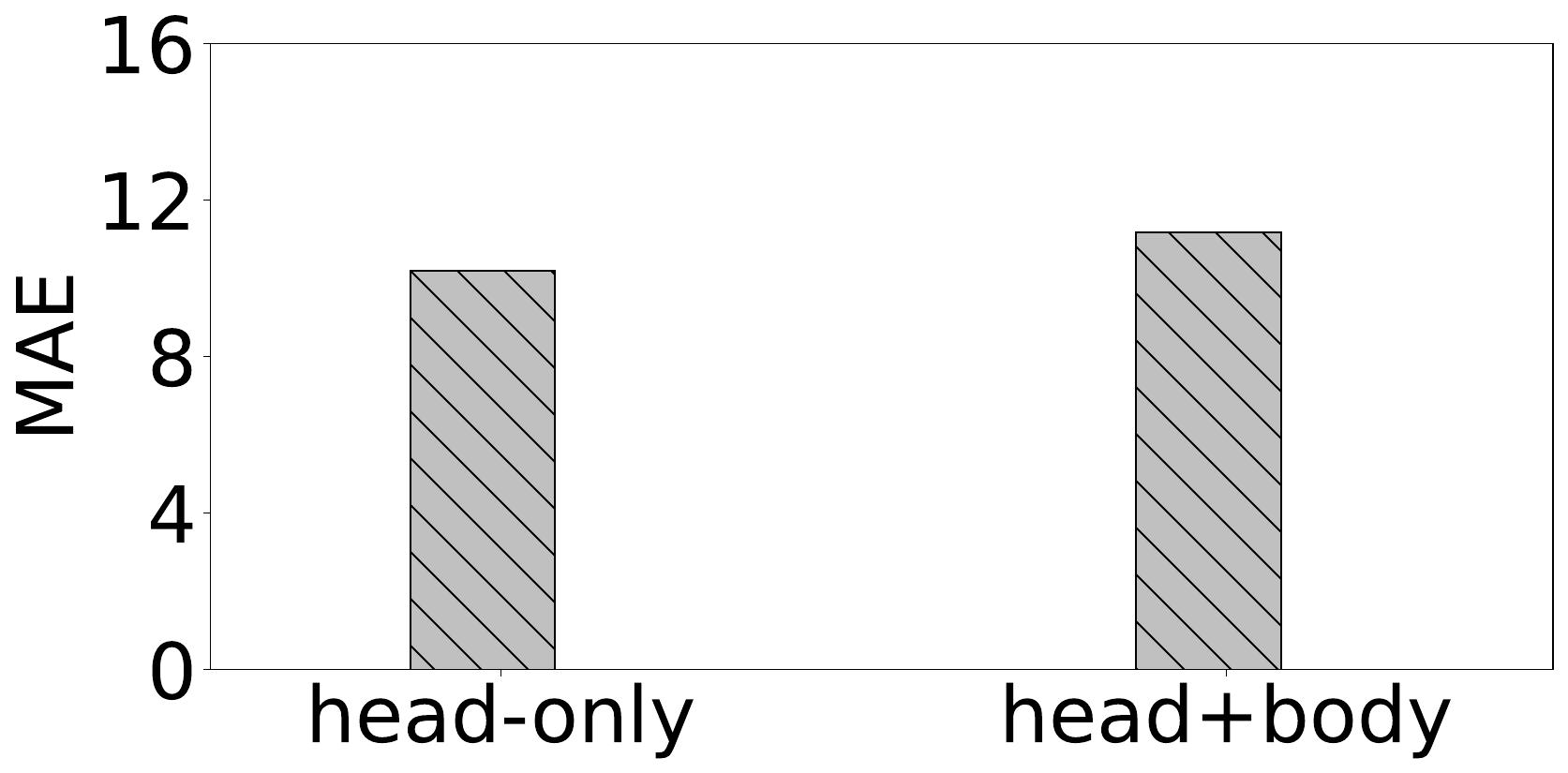}
\vspace{-5pt}
  \caption{Impact of body movements on HME tracking.} 
  \label{fig_exp_headbody}
\end{minipage}
\hspace{.5cm}
\begin{minipage}[t]{0.287\textwidth}\vspace{-0pt}
\centering
\includegraphics[width=.85\columnwidth]{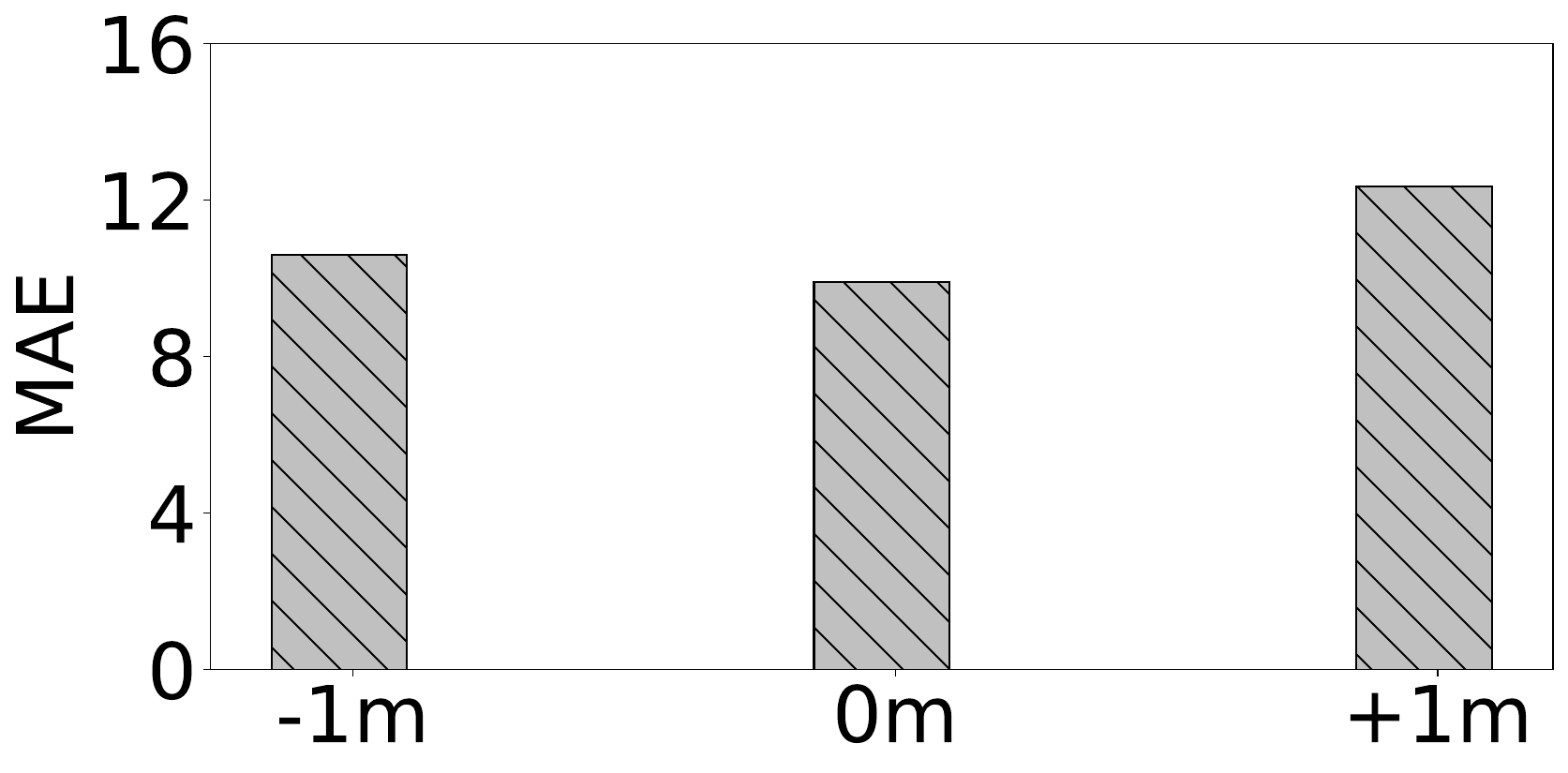}
\vspace{-5pt}
\caption{Estimation errors of the HME at different camera heights.} 
\label{fig_exp_elevations}
\end{minipage}
\end{figure*}

\bheading{HME Effectiveness.} We first validate the effectiveness of the HME because it extracts head movement traces from camera recordings and affects the performance of \sysname. We use Mean Absolute Error (MAE) to measure the absolute angles between the estimated and ground truth head orientation vectors. We focus on the yaw and pitch angles because the roll angle is not needed to locate where the victim is looking at the 360\degrees video. It only indicates how the head tilts (not turns).

\autoref{fig_exp_hpestimator_accuracy} shows the boxplot of MAE (median values marked in red) for different head orientations categorized by their respective yaw and pitch coordinates. The average MAE between estimated and ground truth head orientations are 8.8\degree\ and 4.3\degree\ for the yaw and pitch coordinates, respectively. These errors are comparable to those obtained from generic head estimators for non-HMD limited-range head orientations \cite{ruiz2018fine, kellnhofer2019gaze360,bermejo2020eyeshopper}. Note that the MAE for the yaw coordinate is higher than the pitch coordinate. This is due to the yaw drift effect \cite{feigl2018supervised} where gyroscope sensors incur noises in the yaw coordinate of the ground truth and lead to biases in the yaw estimation. We also observe that the HME performs well in a wide range of yaw and pitch angles supporting 360\degrees video viewing. The performance degrades at extreme angles where fewer samples were collected for model learning. However, since users rarely move to those places, the overall effectiveness of the HME is not affected. 
\looseness=-1

\bheading{Head Tracking on Different Devices.} Next, we test HME performance with two Google Cardboard and Daydream, two popular VR headsets with different shapes, sizes, materials, and colors. We aim to explore how different
HMDs in the camera recording would affect the attack. The average MAE over all head orientations, the MAE of front head orientations looking toward the camera, and the MAE of back orientations turning back away from the camera, are shown in \autoref{fig_exp_headsets}. In general, the MAE of head orientation estimation for Daydream and Google Cardboard are similar, at an average of 10.4\degrees and 9.16\degrees respectively, suggesting \sysname's capability of tracking head movements of victims wearing different HMDs in the recordings.

\bheading{Head Tracking with Body Movements.}
To explore the impacts of body movements, we retrained the HME using two recording cropping methods, one only containing head movements (head-only) and the other containing both body and head movements (head+body). We present the results of MAE in \autoref{fig_exp_headbody}. As shown in the figure, the MAE errors of the head+body setup were 10\% higher than those of the head-only setup since extracting HME-relevant features becomes more challenging when both head and body movements are present. This observation implies background body movements should be cropped out, as used in our default setup. 

\begin{figure}[tb]
\centering
\begin{subfigure}[b]{0.495\columnwidth}
\includegraphics[width=\textwidth]{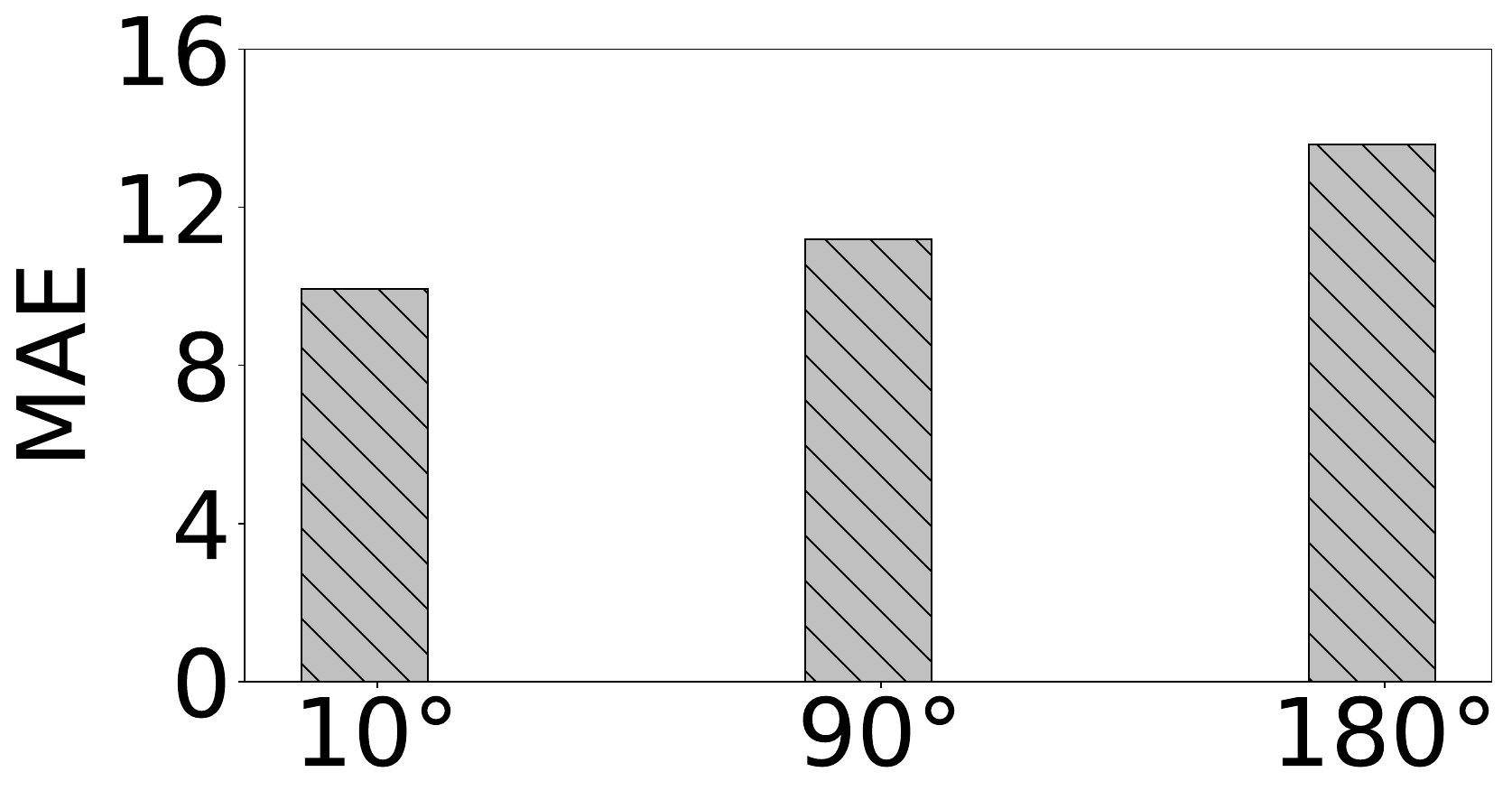}
\caption{Horizontal movements}
\label{fig_exp_deliberate_horizontal}
\end{subfigure}
\hfill
\begin{subfigure}[b]{0.495\columnwidth}
\includegraphics[width=\textwidth]{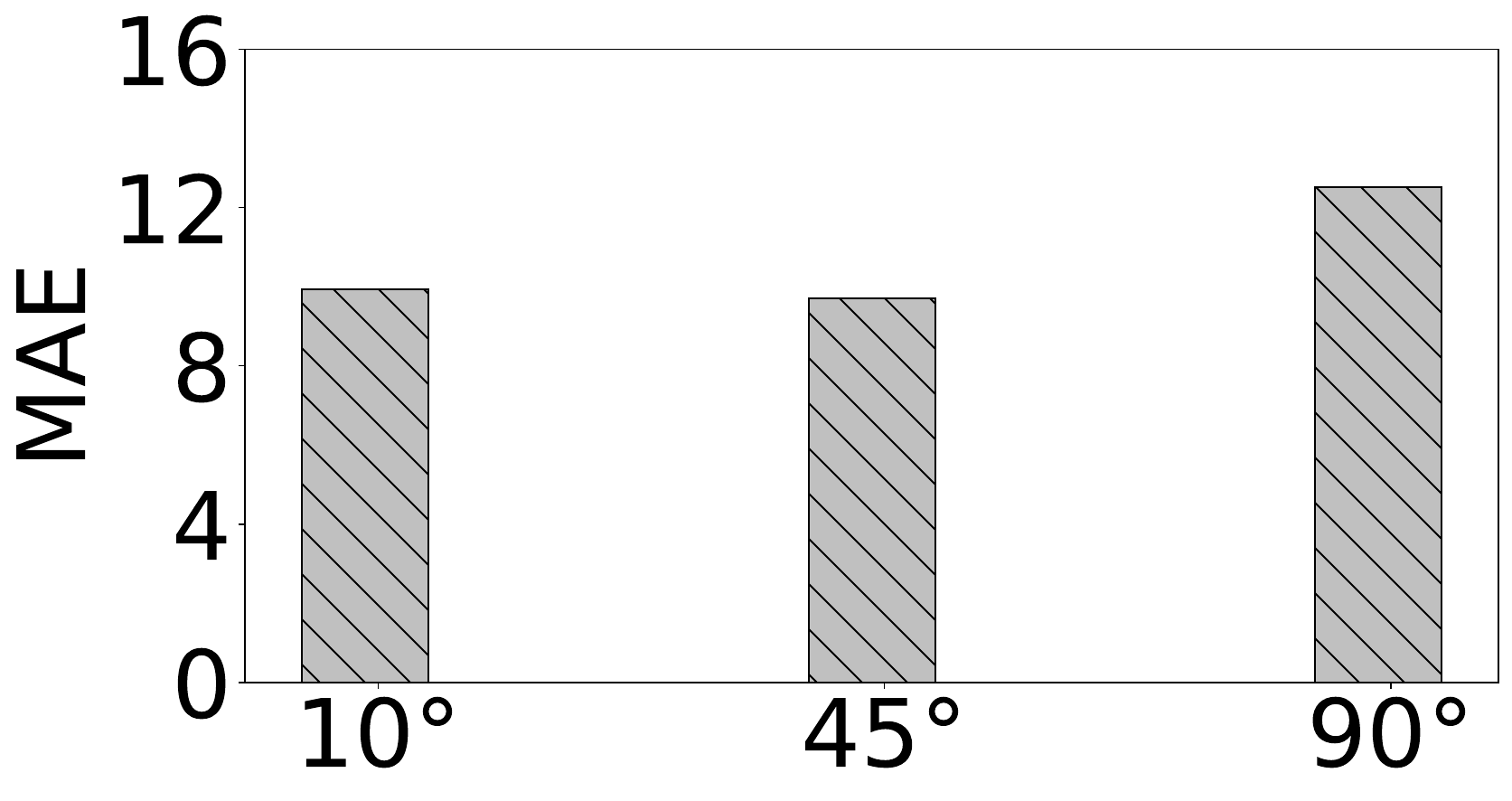}
\caption{Vertical movements}
\label{fig_exp_deliberate_vertical}
\end{subfigure}
\caption{Impact of deliberate head movements on HME tracking.}
\label{fig_MAE_deliberate}
\vspace{-4pt}
\end{figure}

\bheading{Head Tracking with Deliberate Movements.}
Deliberate head movements may disrupt head tracking if they cause the HMD to turn away from the camera, thereby obscuring visual features vital for tracking. We explored the effects of two types of deliberate head movements, vertical angular shifts at 10, 90, and 180 degrees, and horizontal angular shifts at 10, 45, and 90 degrees. For each angle, we collected 20 viewing traces from four users. For each angular shift, users performed five deliberate movements at random moments during a 360 video viewing session and move back to their original positions. The total duration of these movements accounts for about 10\% of a 60-second viewing session without affecting the regular viewing experience. The
average MAE scores are reported in \autoref{fig_MAE_deliberate}. We observe minimal performance degradation for moderate movements, e.g., a vertical shift at 10 or 45 degree and a horizontal shift at 10 or 90 degree, while large deliberate movements at a 180-degree horizontal shift or 90-degree vertical shift decrease tracking performance noticeably. However, these extreme movements rarely occur in practice \cite{dscorbillon} as they require considerable muscle effort and disconnect users from the focused virtual content for a few seconds.

\bheading{Head Tracking At Different Camera Heights.}
We now investigate the performance of HME when the camera is placed at different heights, i.e., the camera being one meter higher than (+1m), at the same height as (0m), and one meter lower than the user’s head (-1m). For each height, we recorded 20 traces collected from four users. The camera was positioned 3 meters away from the user and the average MAE scores are reported in \autoref{fig_exp_elevations}. The MAE errors increase at -1m and +1m because positioning cameras at a height exaggerate the scale of the head movement with respect to the camera. Slightly looking down at 0m appears to look straight down when the camera is placed at +1m. These extreme poses that are challenging to estimate occur more frequently when the camera is placed at -1m and +1m, leading to more errors.

\subsection{Analysis of System Parameters}
\label{sec:eval:freq}

We now proceed to evaluate the identification performance. If one of the top-$k$ inference results is the actual video being viewed, the attack is considered successful. We report the top-$k$ identification accuracy (i.e., attack success rate) across all victim recordings.
In this section, we analyze two system parameters, the recording length and the sampling interval. 

\begin{figure}[tb]
\centering
\begin{subfigure}[b]{0.495\columnwidth}
\includegraphics[width=\textwidth]{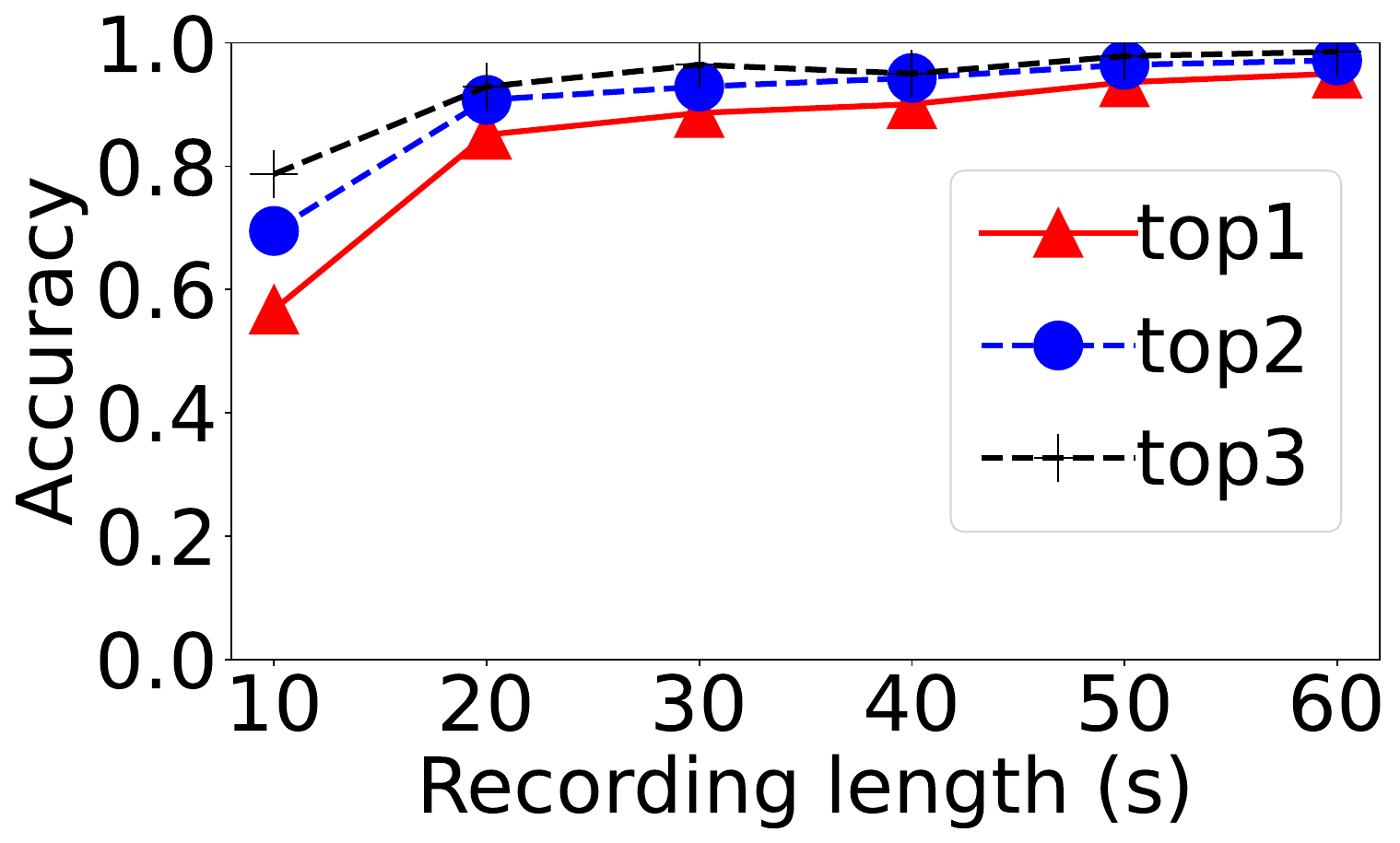}%
\caption{Recording length}
\label{fig_exp_vidpred_duration}
\end{subfigure}
\hfill
\begin{subfigure}[b]{0.495\columnwidth}
\includegraphics[width=\textwidth]{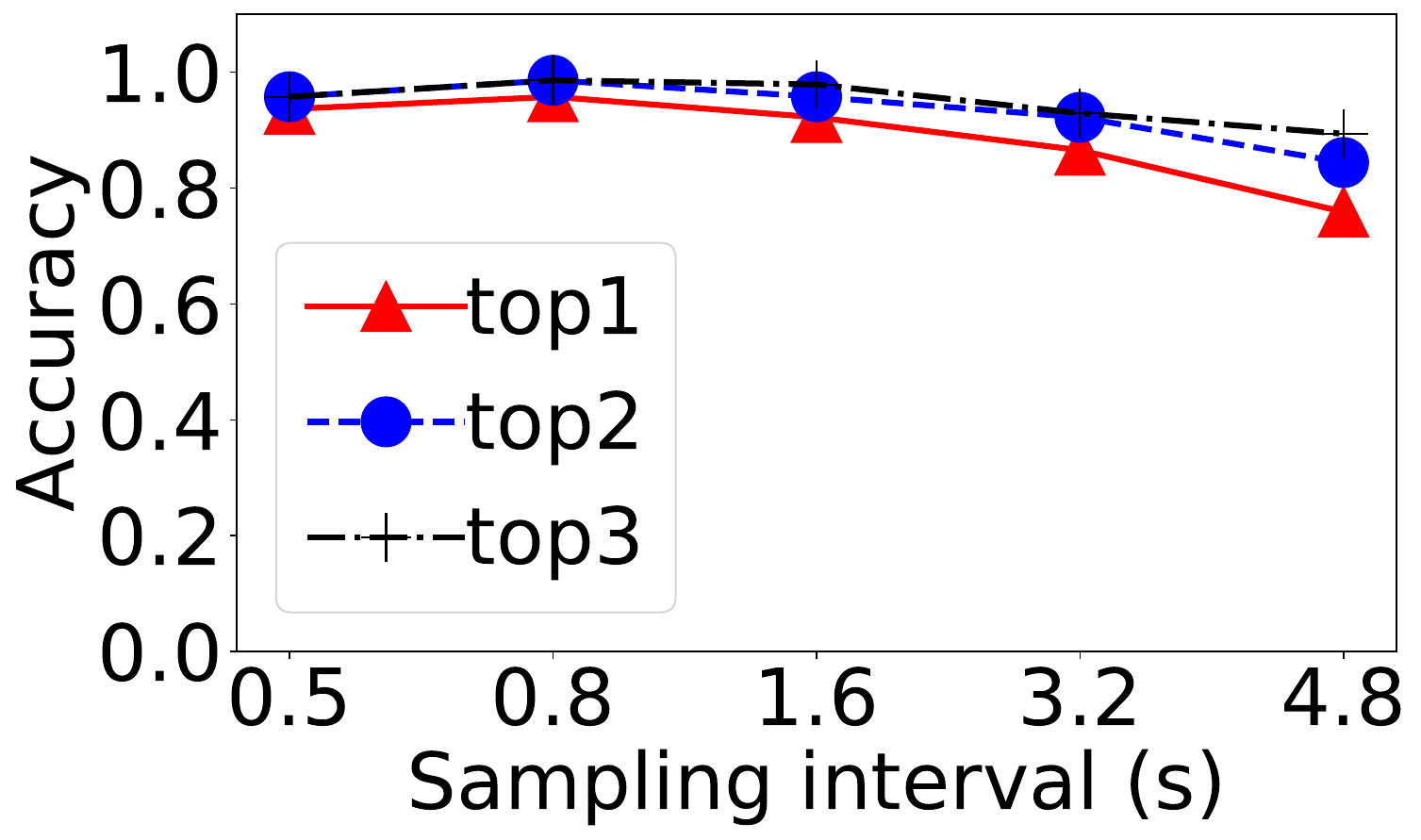}
\caption{Sampling interval}
\label{fig_exp_vidpred_samplingFrequency}
\end{subfigure}
\vspace{-12pt}
\caption{Video identification accuracy versus system parameters.}
\label{fig_system_params}
\vspace{-5pt}
\end{figure}

\bheading{Recording Length.} 
Longer recordings provide more user head movement data and more
clues about the video content being viewed. However, recording the victim for too long increases the risk of exposing the attack. Thus, it is necessary to understand the exact impact of the recording length $T$, the system parameter defined in \autoref{sec:hme:start}. We measure the identification performance when the recording length varies from 10 s to 60 s. We cap the recording length at 1 minute since today's 360\degrees videos typically have several minutes of duration \cite{afzal2017characterization}. The results are shown in \autoref{fig_exp_vidpred_duration}. As expected, the identification accuracy increases as the recording length increases. The top-1,2,3 accuracy of \sysname steadily increases to 96\%, 99\%, and 99\%, respectively, when \mbox{$T=60$}. This result suggests a small trade-off between
the stealthiness and performance of the attack. In case a long recording is not possible, \sysname can utilize a shorter recording, e.g., 20 s, to achieve a lower top-1,2,3 accuracy of 85\%, 91\%, and 93\%, respectively. \looseness=-1

\bheading{Sampling Interval.} 
To minimize the input data while ensuring the attack performance, we study how the sampling interval of the input would affect the top-1,2,3 identification accuracy. We set the $\tau$ parameter discussed in \autoref{sec:identifier:tfm} as 0.5, 0.8, 1.6, 3.2, and 4.8 seconds and repeat the attack to all recordings of victims. As shown in~\autoref{fig_exp_vidpred_samplingFrequency}, \sysname achieves outstanding performance when $\tau$ is 0.5 or 0.8 seconds. For example, the top-1,2,3 accuracy reaches 96\%, 99\%, 99\%, respectively, when $\tau=0.8$. It can also be seen that the accuracy decreases as $\tau$ increases. The reason is that a greater $\tau$ makes the input information sparse, which imposes a penalty on the identification accuracy. On the other hand, minimizing the interval (e.g., 0.5 s) does not necessarily improve the accuracy because it could incur duplicated data samples that do not contribute meaningful information. Thus, we will use $\tau=0.8$ for the rest evaluations.

\subsection{Baseline Comparison} 
\label{sec:eval:baseline}
We also benchmark the performance of \sysname against the following baselines. 
\begin{packeditemize}
    \item {\bf\truth:} 
    Instead of using head movement traces extracted by the HME, the \VideoIdentifier in \truth directly uses the ground truth head movement traces collected by HMD sensors. Since \truth is not impacted by estimation errors of the HME, its performance can be served as the upper bound.
    \item {\bf\hpose:}
    Instead of creating saliency fingerprints, a single head movement trace of a non-victim user is used as the video fingerprint in \hpose. As head movement traces fluctuate significantly and are not ideal for video fingerprinting (see \autoref{sec_mot_how}), \hpose's performance serves as the lower bound.
    \item {\bf\short:}
    To obtain insight into the importance of the design of the \VideoIdentifier, we create \short. It is a variant of \sysname using a reduced version of the trace-fingerprint matching model, where the first two 3D convolutional layers are removed. 
    
\end{packeditemize}

\begin{figure}
    \centering
      \vspace{-10pt}
      \includegraphics[width=0.50\columnwidth]{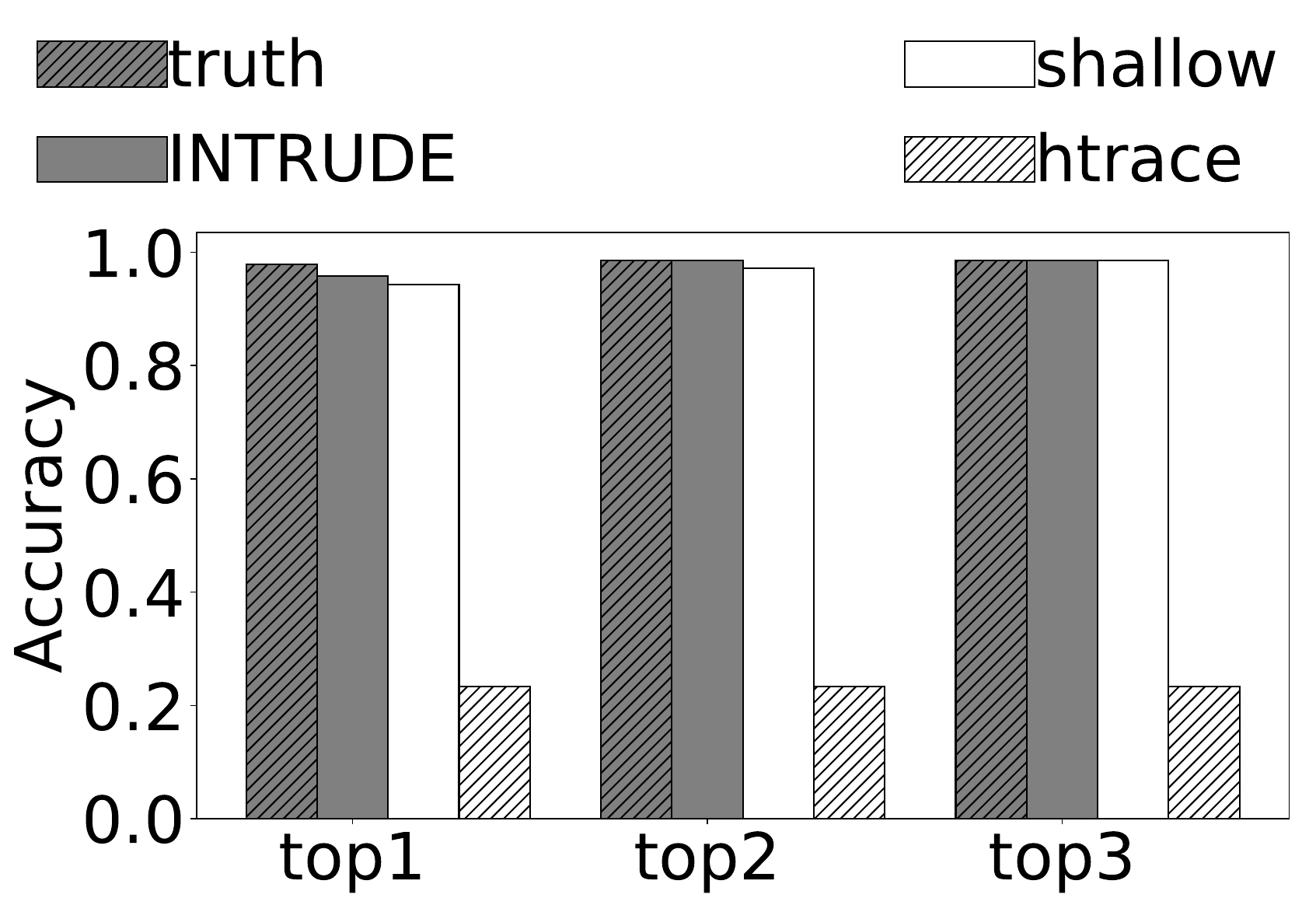}
      \caption{Benchmarking video identification accuracy by comparing \sysname with three baselines.}  
      \label{fig_exp_vidpred_accuracy}
      \vspace{0pt}
\end{figure}

We repeat the experiments in the default setup for each baseline and report the results in~\autoref{fig_exp_vidpred_accuracy}. As expected, \truth has the highest accuracy since it utilizes the ground truth as the input for trace-fingerprint matching and video inference. The performance of \sysname closely approaches the upper bound, achieving a slightly lower yet comparable accuracy to \truth. Fewer convolution layers in \short reduces the top-1 accuracy by 1.4\% because it affects the analysis of the fingerprints and head movement traces and therefore degrades the video identification performance. 
As for \hpose, the identification accuracy is the lowest, indicating that saliency fingerprints are essential features to assist the video matching.

\subsection{Robustness Analysis}
\label{section_eval_robustness}
\label{sec:eval:robustness}
We now evaluate the robustness of \sysname under different environmental factors by repeating the attacks on all the victims. In each robustness study, one environment factor was different from the default setup, e.g., light condition. We captured these additional recordings of victims and launched the attack toward them. Since \truth uses ground-truth head movement traces rather than extracted traces from recordings, it is not affected by the recording environment and thus is excluded in this evaluation. We report the top-1,2,3 accuracy of \sysname and top-1 accuracy of other baselines.
\begin{figure*}[tb]
\centering
\begin{subfigure}[b]{0.246\textwidth}
\includegraphics[width=1.0\textwidth]{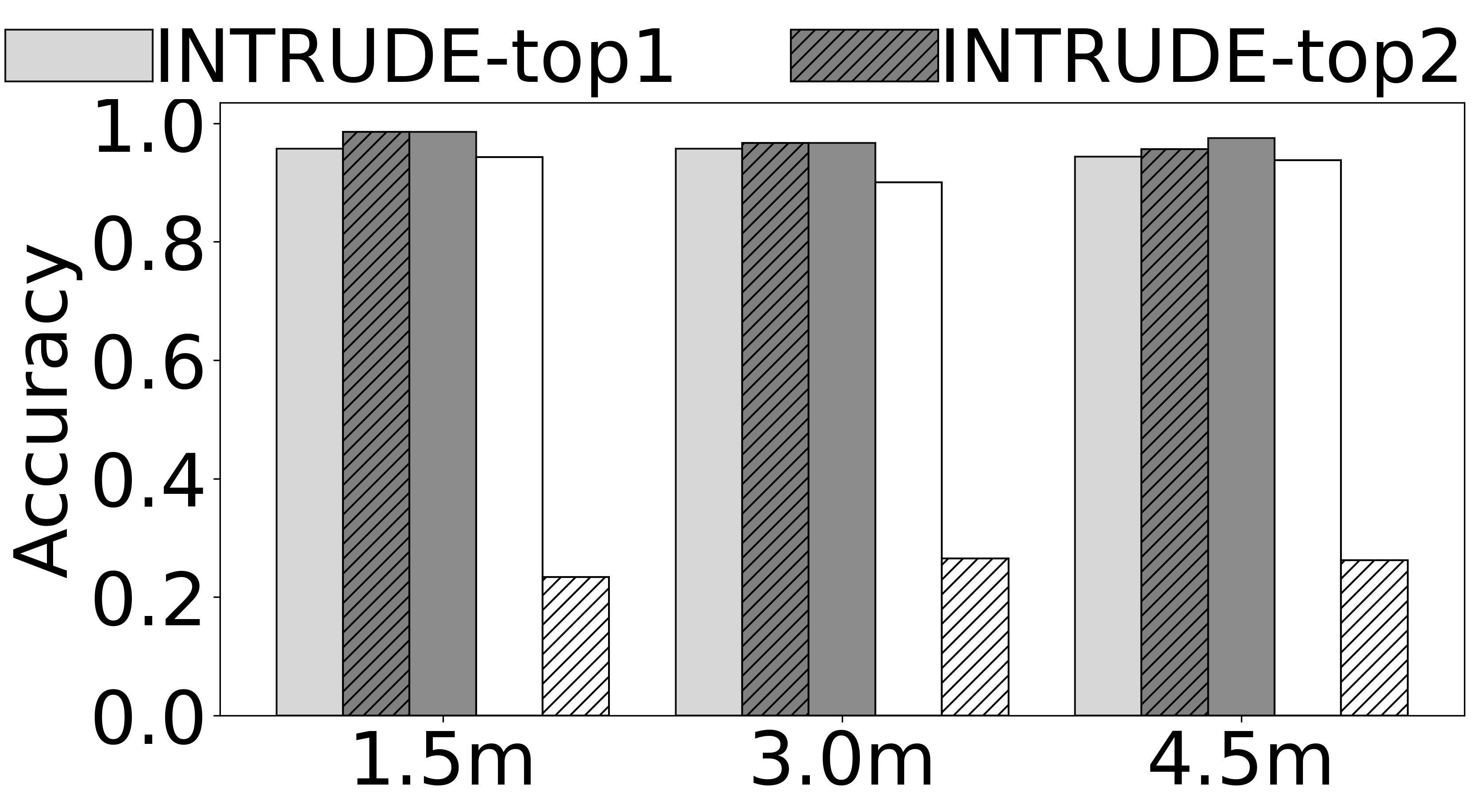}
  \caption{Camera-victim Distance}
  \label{fig_exp_vidpred_distance}
\end{subfigure}
\begin{subfigure}[b]{0.246\textwidth}
  \includegraphics[width=1.0\columnwidth]{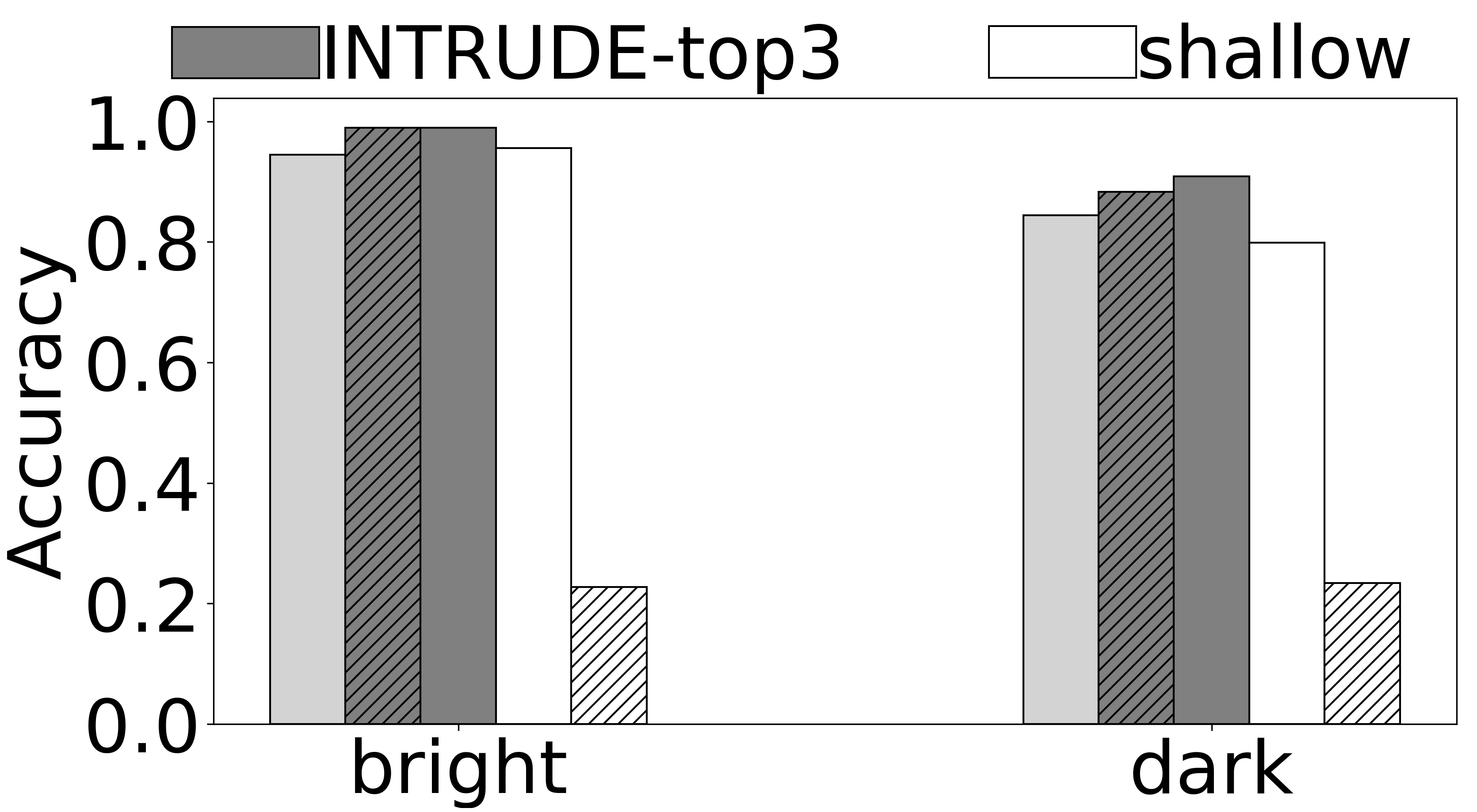}
  \caption{Light Condition} 
  \label{fig_exp_vidpred_lightcondition}
\end{subfigure}
\begin{subfigure}[b]{0.246\textwidth}
\includegraphics[width=1.0\textwidth]{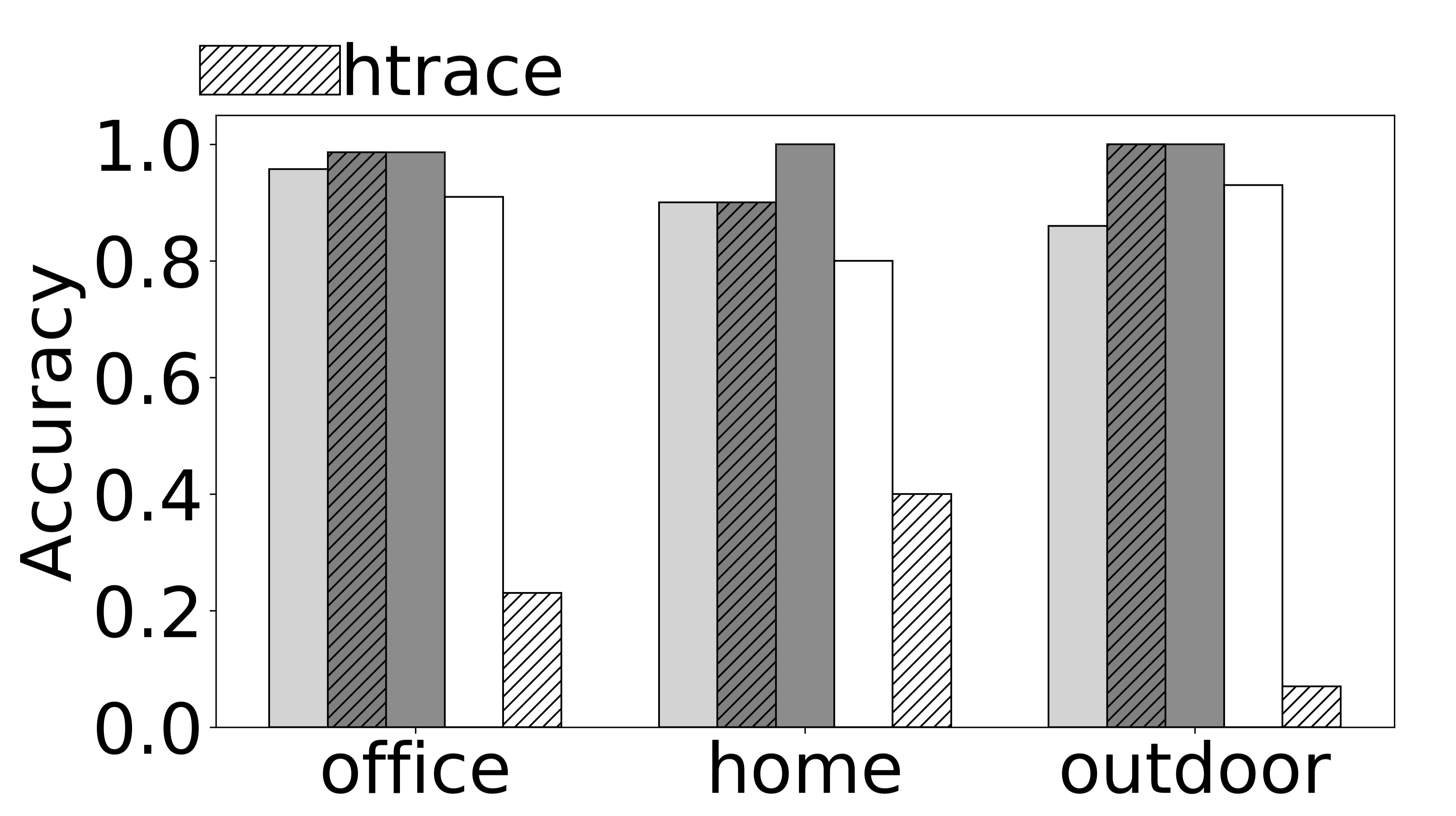}
  \caption{Recording Background}
  \label{fig_exp_vidpred_backgrounds}
\end{subfigure}
\begin{subfigure}[b]{0.246\textwidth}
  \includegraphics[width=\textwidth]{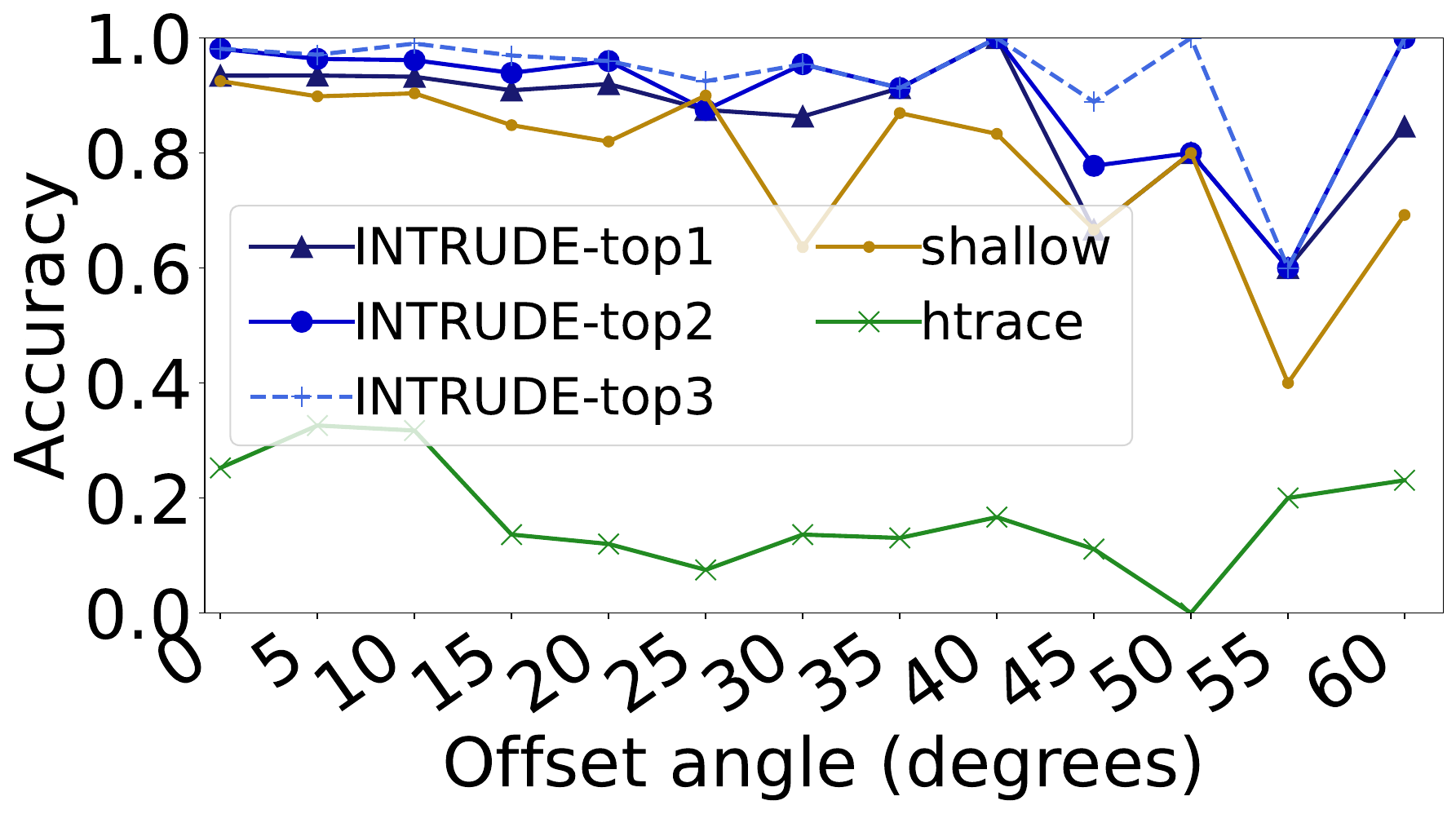}
  \vspace{-6mm}
  \caption{Camera Placement Offset}  
  \label{fig_exp_vidpred_angle}
\end{subfigure}
\vspace{-2.5ex}
\caption{Robustness analysis of \sysname under different environmental factors.}
\label{fig_system_robustness}
\vspace{-2ex}
\end{figure*}

\bheading{Distances.} 
To measure the impact of the distance between the victim and the recording camera, we place the camera at 1.5 m, 3.0 m, and 4.5 m, away from the victim, respectively. 
The results in~\autoref{fig_exp_vidpred_distance} show that the performance of \sysname is stable as the distance varies. The attack distance only affects the clarity and visual details of the recordings and the extracted head movement trace. As long as the number of captured pixels for the head is sufficient, a long distance does not affect the identification accuracy. The distances in our evaluations are greater than or equal to those in prior recording-based attacks \cite{ling2019know,ye2017cracking}. To further extend the distance, we can utilize an advanced camera to capture 2K or 4K videos. The accuracy of \short decreases a little as the distance increases because it is a weaker model with a low capacity to learn spatiotemporal patterns in the head movements and video fingerprints.

\bheading{Light Conditions.} 
The proposed attack may be launched in various light conditions depending on when the attacker finds the victim vulnerable. If the camera is placed in a dim environment, the recording could be noisy and might contain indiscernible regions~\cite{chen2018learning}.
Extracted head movement traces can then become erroneous, eventually affecting the attack performance. 
We launch \sysname on recordings captured in the bright indoor condition ($\sim$1000 lux) and the dark indoor condition ($\sim$50 lux) according to the lighting definitions in \cite{gsa2020}. The bright condition is equivalent to a lit office while the dark condition is similar to a dark corridor or an unlit parking lot. 
As shown in \autoref{fig_exp_vidpred_lightcondition}, there is a small performance drop due to the effect of the dim environment. 
The top-1 accuracy of \sysname decreases from 96\% to 84\% while the top-1 accuracy of \short drops from 92\% down to 79\%. However, we believe \sysname still poses a threat since the top-2,3 accuracy in the dim environment can reach 90\% and 93\%.

\bheading{Background Variations.} 
The location of the attack affects the background scenes in the recording. These visual backgrounds of the victim are analyzed to estimate head orientations and thus have an impact on the performance of \sysname. 
Therefore, we further investigate \sysname in different background scenes. 
In addition to the open office used in the default setup that includes gray cubicles and blue walls, we conduct the attack in two additional backgrounds.
One is the home living room where the background scene consists of wood furniture, brown doors, and dark orange walls. The other is the outdoor background that is a typical garden with lawns and plants.
Within each background category, we make the exact scene different for each victim to ensure the diversity of the recordings under evaluation. For example, different plants may appear behind a subject in the outdoor case.   
The results are shown in~\autoref{fig_exp_vidpred_backgrounds}. Despite the slight variation, \sysname achieves a satisfactory performance across different backgrounds. The average top-1,2,3 accuracy across three backgrounds is 91\%, 96\%, and 100\%, respectively. The support of different backgrounds stems from the HME design. As illustrated in~\autoref{sec_headpose_estimator}, the HME learns to focus on the HMD device and ignore the background noises. \looseness=-1

\bheading{Camera Placement Offset.}
In the default setup, the camera is placed in front of the user along the user's front-facing direction, i.e., there is no offset angle between the filming direction and the front-facing direction (see \autoref{sec_convert_headoren}). In this section, we place the camera in different locations to vary the offset angle between the filming direction and the head's front-facing direction. When the offset angle increases, the camera is gradually moved to the side of the victim. The results in \autoref{fig_exp_vidpred_angle} show that there is a slight decrease of top-1 accuracy when the offset angle increases, e.g., around 85\% for \sysname at the offset angle of 60\degree. This is due to the increased errors in head orientation estimation when the face and the HMD rotate away from the filming direction (see \autoref{sec:eval:hme}). Despite the decrease in top-1 accuracy, we note that \sysname still performs well, with a top-2 and top-3 accuracy at 100\%, at the offset angle of 60\degree. We also observe a bit of fluctuation when the offset angle is large. This is because head orientation estimation becomes less stable in these cases. More evaluation data can be used to smooth out the variance.
\looseness=-1

\begin{figure}
    \centering
    \includegraphics[width=0.55\columnwidth]{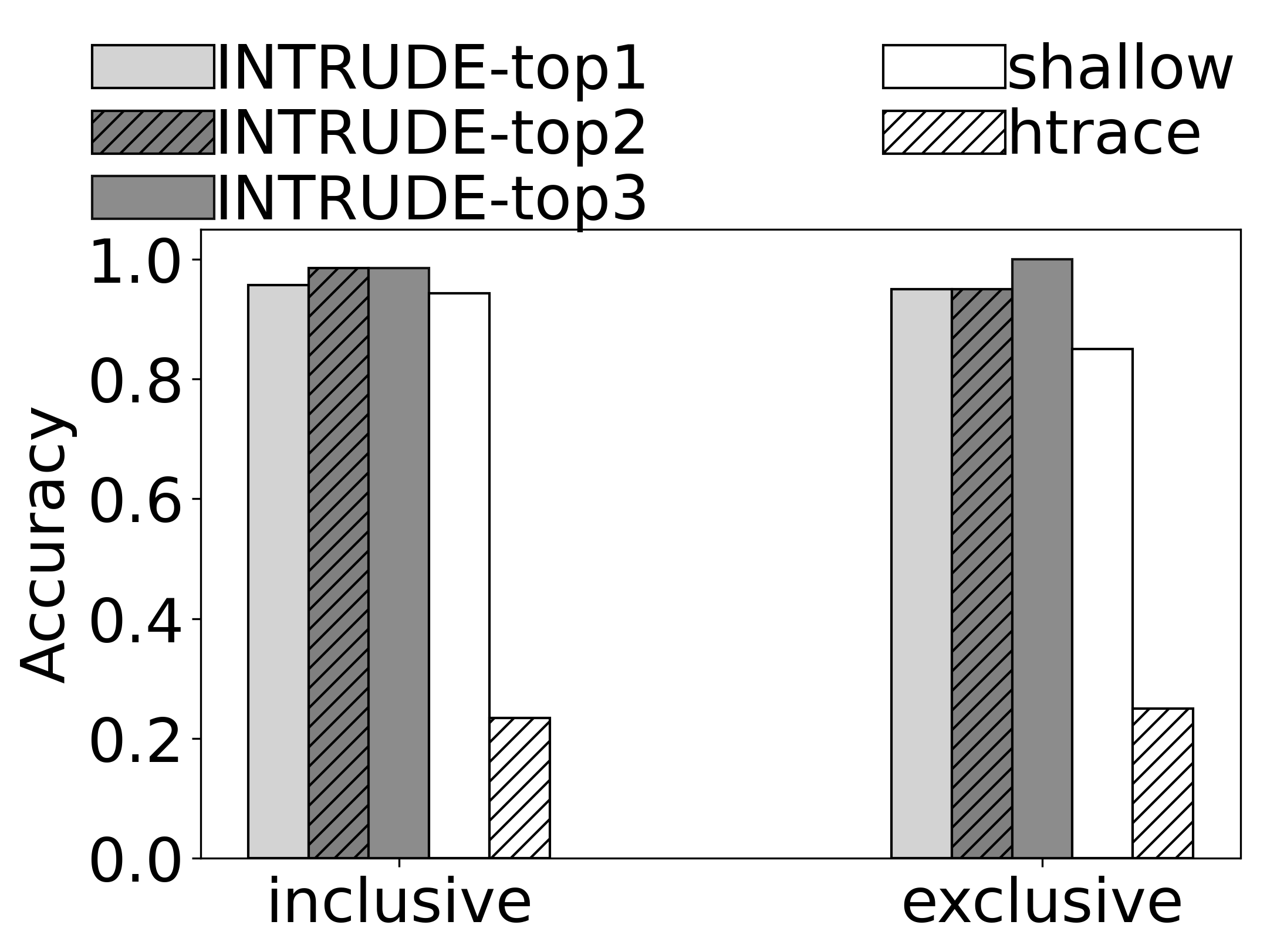}
      \caption{Identification accuracy of \sysname when additional participants are used exclusively as victims.}  
      \label{fig_exp_vidpred_exclusive}
      \vspace{-1.5ex}
\end{figure}

\bheading{Participants Who Are Not in the Training Set.} 
As mentioned in~\autoref{sec:exp:settings},
the participants in the above experiments took part in both tasks (training the HMD and playing the role of the victim). To evaluate how the knowledge about the procedures of the former task influences the results of the latter task, we additionally recruited four new participants who exclusively played the victim role and trained the HME only via data from the existing 31 participants. As shown in \autoref{fig_exp_vidpred_exclusive},
\sysname achieves top-1, 2, 3 accuracies of 95\%, 95\%, and 100\% when the participants are exclusively the victims. The performance is comparable to the inclusive case where the participants appeared in both tasks. This indicates involving participants in both tasks does not affect \sysname's performance, validating the appropriateness of our experiment methodology.

\subsection{Open-World Video Identification}
\label{sec:eval:open}

We have evaluated video identification in the in-library scenario, where the video viewed by the victim is known to be from the library.  This scenario has been assumed in the majority of prior video identification attacks \cite{reed2016leaky,gu2019traffic,maiti2019light,xu2014watching}. However, another scenario exists in practice -- open-world video identification, where we do not know if the video being viewed is in the library. The accuracy of open-world identification has been evaluated by theoretically estimating the accuracy drop using a statistical model \cite{schuster2017beauty}. We use this approach to evaluate the degradation of \sysname in the open-world scenario as the number of out-of-library videos increases. 

To model the accuracy drop statistically, 
the Bayesian Detection Rate (BDR) can be used. BDR estimates the ratio of successful identification among all identification attempts and can be formally expressed as, 

\vspace{-6pt}
\begin{equation}
    \text{BDR} = \frac{\text{TPR} \times {base}}{\text{TPR} \times base + \text{FPR} \times (1 - {base})} 
\end{equation}
\vspace{-6pt}

where TPR (True Positive Rate) is the top-1 accuracy of in-library identification reported above. FPR (False Positive Rate) can be derived from TPR via $FPR = (1-TPR)\frac{P}{N}$, where $N$ and $P$ are the numbers of negative/positive matches (\autoref{sec_offline_training}) \sysname encounters during in-library identification, respectively. The base rate ($base$) is the probability of a test video being in-library, i.e.,

\vspace{-10pt}
\begin{equation}
    base = \frac{\text{Number of in-lib videos}}{ \text{Number of both in-lib and out-of-lib videos}}
\end{equation}
\vspace{-10pt}

Based on our evaluation results, \sysname achieves TPR of 0.96 and FPR of 0.000068. Given $base=0.001$, the open-world dataset (in-library plus out-of-library videos) is 1,000 times larger than the target library, totaling 635,000 videos. In this case, the BDR is 93\%. Given $base=0.00025$ when the open-world dataset expands to 2,540,000 videos, the BDR achieves 78\%. Our analysis indicates \sysname still poses great threats even when there are several million 360\degrees videos in the open world.

\section{Discussion}
\subsection{Limitation and Potential Improvement}
\bheading{Re-training Cost.} Launching \sysname\ on a new HMD requires the attackers to offline collect recordings of the target HMD and retrain the head orientation estimation model for \sysname\ to recognize the HMD's shape and color. The attackers can capture the recordings of their own head movements when using the target HMD. The brand and model of the HMD can be visually recognized when targeting the victim. Despite the extra model training, the retraining cost is not prohibitive. First, there are only a handful of HMDs on the market. We only need to repeat the training a few times based on demand to make \sysname\ generally applicable. Second, the retraining data can be collected by fewer subjects using the HMD in different outfits. Along with the data augmentation we used (\autoref{sec_offline_training}) and generating additional computer-synthesized data  \cite{bermejo2020eyeshopper}, the retraining cost can be minimized.\looseness=-1

Notably, the trace-fingerprint matching model of \sysname is not HMD-dependent. The current model can be directly used and no retraining is needed when applying \sysname to a new HMD. The reason is that the trace-fingerprint matching model takes extracted head movement traces and video fingerprints as input rather than the recording of the new HMD. The TFM can thus infer the video titles in an HMD-agnostic way.\looseness=-1

\bheading{Non-continuous Video Playback.}
We have so far considered continuously-playing videos. This is the most common scenario for 360\degrees video viewing in HMDs because, unlike 2D video viewing with control shortcuts on keyboards, users tend to continuously immerse themselves in 360\degrees video scenes without interruption \cite{oh2019effects}. However, scenarios where a user pauses, rewinds, or fast-forwards the video can occur, disrupting the continuity of head movements and potentially affecting the attack's success. Fortunately, these infrequent instances can be identified due to its distinctive interactive sequence. For example, pausing necessitates two clicks (one to pause, another to resume). Fast-forwarding or rewinding may require either a single click on the progress bar to skip content or three clicks (pause, forwarding/rewinding button, resume).  An alignment module can then be added to \sysname to process the recording such that the victim's head movement is aligned with the continuous video playback. In the case of a pause, this new module would discard the recording period while the video is paused. For rewind or fast-forward actions, it extracts the head movement trace once the playback resumes and temporally shifts this trace forward or backward to a video-playing position where the revised trace matches the saliency fingerprints with the highest confidence scores. 
Similarly, this alignment module can also be used to align the video's start time and the moment when the user clicks the play button if there is a small delay in between.

\subsection{Mitigations}
\sysname is effective because video information is always leaking as long as attackers can observe the victim. An approach to prevent the attack is eliminating the direct line of sight between the attacker and the victim \cite{long2011no}. Users could find a large non-transparent obstacle between them and the crowds or turn their heads away from directions that may expose their head movements. However, these efforts are burdensome for casual VR usage and may not always be feasible in certain environments. Moreover, this prevention approach depends on the user's judgment and vigilance of the surrounding people. It is unlikely to work for advanced attackers who disguise themselves or utilize miniature spy cameras \cite{wu2014security}.

Besides prevention, one could mitigate the success rate of the attack by complicating the attack. For example, in \sysname, the roll axis of the VR coordinate system is aligned with the victim's head orientation at the time when the video starts playing. The attacker exploits this fact to extract the victim's head orientation vector from the recording. To undermine the HME, the VR engine can randomly set the direction of the VR roll axis when the video starts. However, this approach cannot completely eliminate the risk,  as the victim will eventually follow a certain head movement pattern influenced by the evolving salient content. The attacker may still identify the correct VR roll axis by spatially shifting the saliency maps to positions with the highest confidence scores. More importantly, users will feel disoriented and have motion sickness every time a video starts playing because the VR coordinate system is not aligned with the head orientation. A full-scale study of the defense capability and usability is needed before this mitigation can be applied.

\section{Related Work}
\label{sec:related}
\bheading{Recording-based Side-channel Attacks.}
Capturing recordings through the visual channel has been used for recovering keystrokes and unlock patterns on mobile devices such as smartphones and tablets. Existing works have recorded victims (including body movements \cite{khan2018evaluating, li2005association, chen2022armspy}, eyes \cite{sun2016visible,chen2018eyetell}, fingertips \cite{ye2018video,balzarotti2008clearshot,yue2014blind,shukla2014beware}), reflective surfaces~\cite{backes2008compromising,backes2009tempest,xu2013seeing6,raguram2011ispy5}, and backside motion of tablets~\cite{sun2016visible} to perform inference attacks. Recently, there are works showing that sensitive information such as keystrokes entered by participants can be retrieved from virtual  backgrounds~\cite{hilgefort2021spying,sabra2022background}, or reflections from participants’ glasses \cite{wasswa2022proof,long2022private}.

There is one existing work~\cite{ling2019know} exploring recording-based side-channel attacks to infer the passwords or keystrokes from users' head movements in HMDs. The attack used optical flow techniques to track the rotation movement of the head in the recording. However, it can only be used for analyzing a very short duration (\ie, during keystrokes) and limited rotation range (\ie, within the virtual keyboard), because this technique has low accuracy and is susceptible to noises. Therefore, they cannot be applied to the case of 360\degrees video viewing, where we must continuously track the head orientation (omnidirectional) over a long time (up to 60s in our case). 
In this paper, we utilize recordings of head movements of HMD users to launch 360\degrees video identification attacks. Instead of harnessing an unsupervised hand-crafted technique used in~\cite{ling2019know}, we leverage deep learning techniques to train the head orientation estimation model. This enables \sysname to achieve state-of-the-art estimation performance while being able to track the user's head movements in 360\degrees video viewing, where the user may look in any direction. 
To the best of our knowledge, we are the first to use the recording-based visual channel to identify 360\degrees videos in VR HMDs.
\bheading{Side-channel Attacks on HMDs.}
Due to the increasing popularity of VR/AR devices, side-channel attacks on HMDs are gaining more attention. To date, existing side-channel attacks on VR/AR devices~\cite{luooculock,luo2022holologger,al2021vr,ling2019know,meteriz2022keylogging, zhang2023s, slocumgoing, wu2023privacy,heimdall} mainly aim to infer the keystrokes when the user is wearing an HMD. 
There is one recent work~\cite{slocumgoing} utilizes head motions to infer keystrokes in AR/VR devices,
but it relies on a pre-installed malicious app to record sensor readings directly.
VR-Spy~\cite{al2021vr} utilizes channel state information of WiFi signals to recognize keystrokes in VR headsets.
Similarly, keystroke inference attacks have been performed on AR devices, e.g., Microsoft Hololens~\cite{meteriz2022keylogging,luo2022holologger}. 
Meteriz~\etal~\cite{meteriz2022keylogging} explore the AR hand tracking features in Microsoft Hololens, while Luo~\etal~\cite{luo2022holologger} use the six degree-of-freedom motion tracking information.  
Unlike these works, the goal of \sysname is not to infer the keystrokes, but rather the 360\degrees video being viewed in the HMD.\looseness=-1

\bheading{Side-channel Attacks for Video Identification.}
Among different side-channel techniques to perform video identification attacks, traffic analysis is the most popular type due to the popularity of online video streaming platforms such as Netflix and the ease of accessing the encrypted network traffic in wireless networks. To date, most of the existing works focus on inferring traditional 2D videos~\cite{reed2016leaky,schuster2017beauty,gu2019traffic,song2020vtim,saponas2007devices,liu2008wavelet,liu2010video,reed2017identifying,dubin2017know,li2018deep}. Recently, Bae~\etal~\cite{baewatching} show that video identification attack is feasible in Long Term Evolution (LTE) networks by utilizing broadcast radio signals. 
Other side channels
such as powerline electromagnetic emancipation~\cite{enev2011televisions}, reflective lights on windows~\cite{xu2014watching}, power measurements at the smartphone charging hub~\cite{acharya2021phone}, and luminance of smart light bulbs~\cite{maiti2019light} have also been utilized to identify 2D videos.

As for 360\degree\ videos, a recent study~\cite{kattadige2021360norvic} distinguishes 360\degrees videos from regular videos (\ie, a binary classification) for mobile network providers by using packet- and flow-level traffic of encrypted streams.  
By contrast, \sysname is a side-channel attack that identifies 360\degree\ video tiles while they are being viewed in HMDs. We harness a novel side channel, \ie, the head movement, which does not exist when viewing conventional 2D videos on laptops and smartphones.

\section{Conclusion}
In this paper, we investigate the leakage of 360\degrees video titles from VR HMD users. In contrast to the public impression that HMDs conceal the displayed content, we show that \sysname can infer video titles with high accuracy by recording victims whose vision is blocked and leveraging the unique side channel of head movements. In a typical indoor setup with the camera a few meters away from the victim, \sysname can identify video titles with top-1,2,3 accuracy of 96\%, 99\%, and 99\%, respectively. Furthermore, \sysname performs well under various noises in different recording distances, lighting, background scenes, and camera placement angles, as well as in the open-world identification scenario with out-of-library videos. \looseness=-1

\section{Acknowledgement}
This work was in part supported by National Science Foundation grants NSF-IIS 2140620 and OAC 2144764

\newpage

{
\let\oldbibliography\thebibliography
\renewcommand{\thebibliography}[1]{%
  \oldbibliography{#1}%
  \setlength{\itemsep}{1pt}%
}

{
\small
\bibliographystyle{IEEEtranS}
\bibliography{}
}

\appendix

\section{Appendix}
To accurately track the head movements of victims in camera recordings, it is crucial for the Head Movement Estimation (HME) scheme to precisely estimate head orientations at a granular level. This requires the HME training set to include all possible head orientations that victims may exhibit. 
In this section, we explain the comprehensive collection process for all common head orientation angles for HME training. We begin with a basic explanation of the VR coordinate system, necessary for understanding the data labeling procedure, followed by a detailed description of the experimental setups for collecting training set samples. Finally, we outline the annotation process required to generate the ground truth.

\subsection{Specifications of Coordinates in Virtual Reality}
The VR OS tracks user head movements using the VR coordinate system, a critical factor in rendering objects in VR. 
Every head movement triggers the system to adjust the positions of virtual objects relative to the viewport, thereby creating the illusion of user exploration within the virtual environment. The system represents head orientation as a rotational operation. This operation rotates a unit vector, i.e., a vector where one component equals 1 while the remaining components equal zero, from the reference axis (the roll axis in \autoref{fig_hmdcoord}) to the current orientation of the head. Many VR engines use quaternions, a tuple of 4 elements equivalent to a $3 \times 3$ rotation matrix, to represent the rotation operation. The quaternion is preferred as the rotation matrix due to its compactness \cite{hanson2005visualizing}. As such, the 3D head orientation vector can be derived simply by applying the rotational operation to the unit vector from the reference axis.

\subsection{Experiment Setup}\label{appendix:exp}
This section discusses the setup and procedure to collect samples for the training set. The data collection procedure includes three main steps, designing a virtual room, designing a viewing procedure, and capturing camera recordings.

\subsubsection{Design of the Virtual Room} We design a virtual room that enables subjects to exhibit all aspects of their head and HMD movements to the recording camera (\RC).
The virtual room consists of five cubes placed around the subject, including two colored cubes and three gray cubes. 
The cubes' arrangement is depicted in \autoref{experiment_phase1}.  All cubes are initially suspended 4.5 meters above the floor and descend at a constant speed. After 30 seconds, the cubes reach the floor and subsequently disappear. \autoref{experiment_phase1_room} shows part of the virtual room with 3 gray cubes visible in the viewport.

\begin{figure}[!t]
\centering
\subfloat[] {\includegraphics[width=0.21\textwidth]{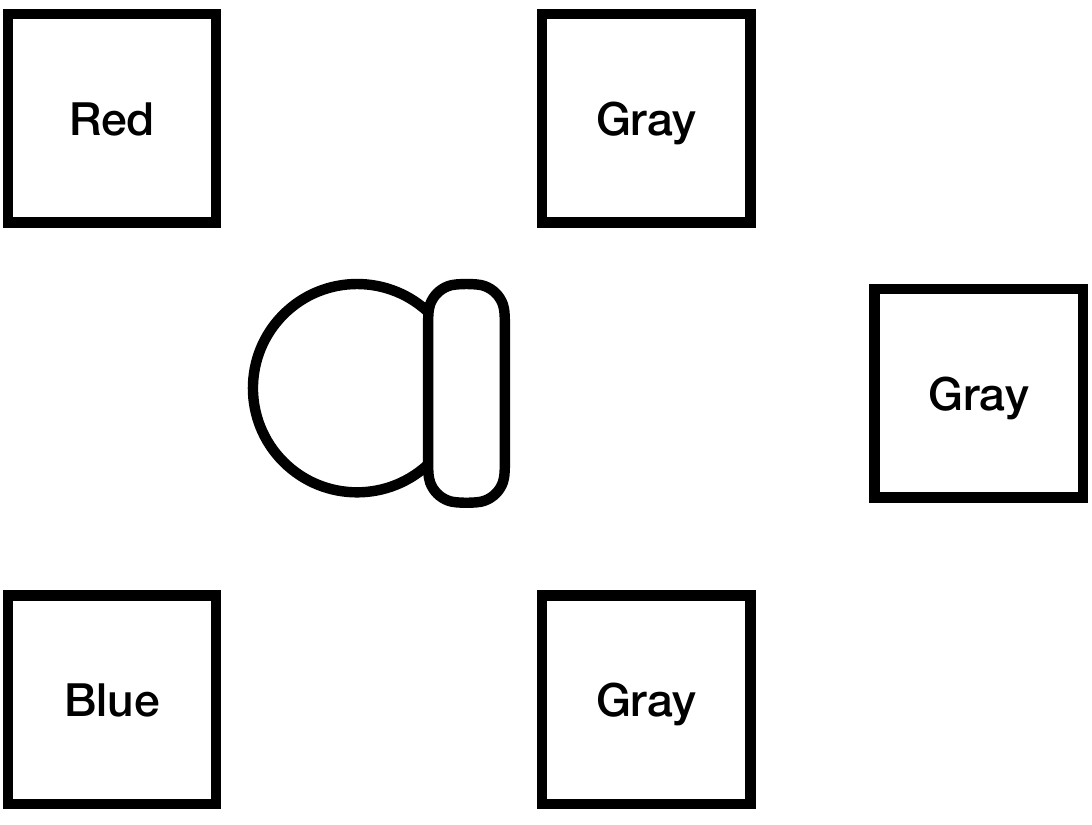}
\label{experiment_phase1_design}}
\hfill
\subfloat[]{\includegraphics[width=0.25\textwidth]{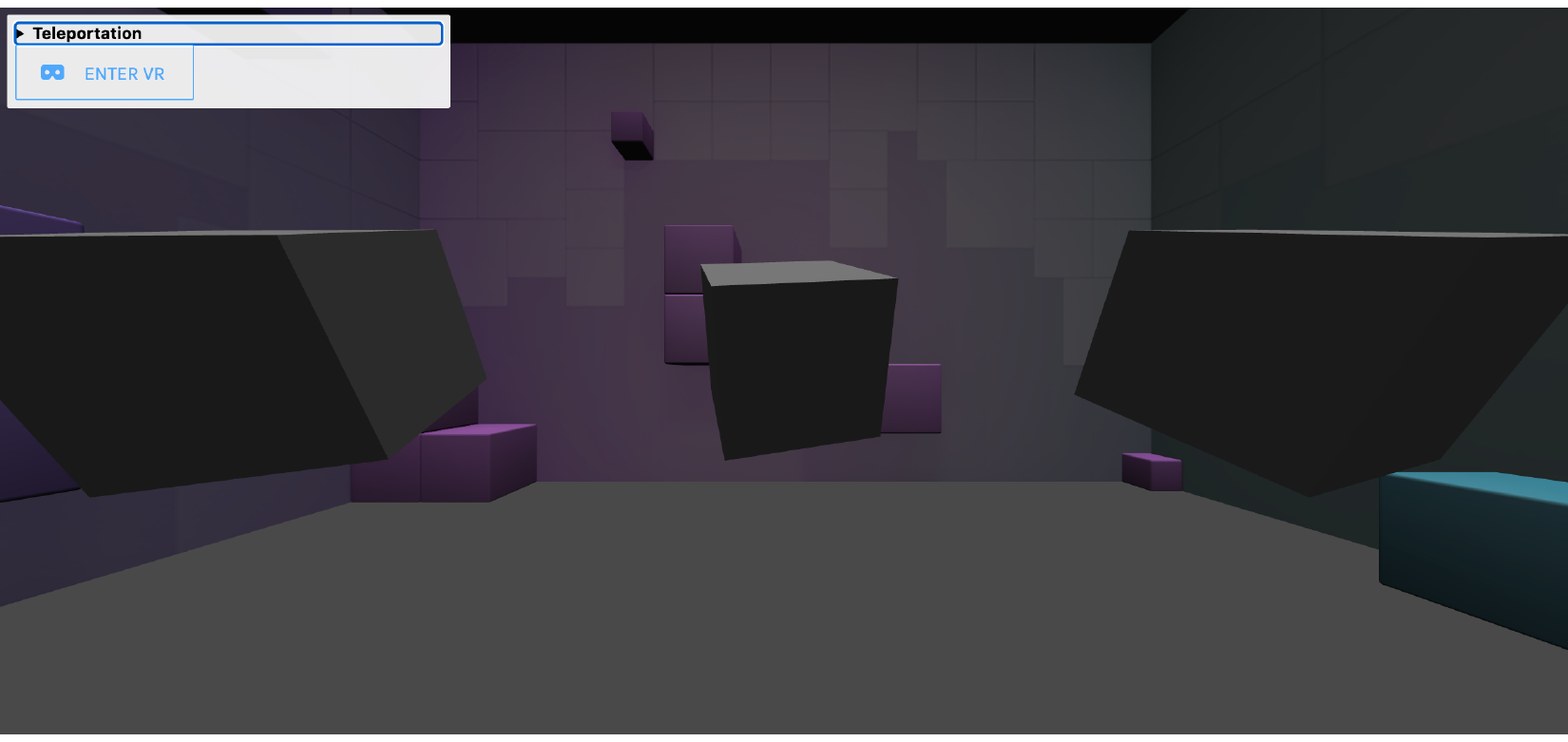}
\label{experiment_phase1_room}}\\
\caption{Design of the virtual room. a) the positions of cubes around a subject b) a view inside the virtual room.}
\label{experiment_phase1}
\end{figure}

\subsubsection{Viewing Procedure}
After putting on the HMD, subjects are asked to look around in a systematic way such that \RC could capture their heads at different angles. In particular, subjects turn their heads from the red cube, glance over the gray cubes, and stop briefly at the blue cube before starting over from the blue cube to the red cube.
As the head follows the dropping cubes, the \RC placed at a distance away captures their head poses.
After 30 seconds, the cube drops to the floor, subjects are told to look around and explore locations in the room that have not been seen in the first 30 seconds. This procedure lasts for 60 seconds. \looseness=-1

\subsubsection{Capturing the Camera \Observations}
We used the same set of subjects and followed the same setup described in \autoref{section_setup} to capture camera recordings. The virtual room was installed in the HMD. Three \RCs were positioned in front of the participants at distances of 1.5m, 3.0m, and 4.5m. Participants were instructed to wear the HMD and follow the pre-established viewing procedure while the \RC captured the camera recordings. Simultaneously, the timestamped head movement data was collected from the HMD gyroscope sensors and stored in head movement logs. These logs will be used for subsequent annotation. This procedure resulted in camera recordings capturing head poses of subjects ranging from $-179\degrees$ to $179\degrees$ in yaw and from $-55\degrees$ to $55\degrees$ in pitch.

\subsection{Labeling the Camera \Observations} 
The camera recordings were divided into individual frames, with each frame's head orientation annotated. Following this, we identified and removed a source of noise from the annotated data.

\subsubsection{Data Annotation}
Our method for annotating head poses in each frame of the camera recordings is both economical and precise, thanks to the availability of the gyroscope sensors within the HMD. To generate a large volume of annotated data, the recordings and the gyroscope data from the head movement logs were synchronized. Upon synchronization, every frame was correctly linked to a head orientation vector from the head movement log. The primary challenge was determining the time offset between the camera recording and the head movement log. Correctly calculating this offset requires the annotator to manually identify a key frame $f_i$ at time $t_i$ from the camera recording and the associated key head orientation $v_j$ at time $t_j$ in the head movement log.
Let $((f_i, t_i)$, $(v_j, t_j))$ denote the selected pair before the synchronization.
Then, the offset is the time interval between the two key points, $t_i - t_j$. After being synchronized, $(f_i, v_j, t_j)$ is the annotated frame with the ground truth head orientation vector and the adjusted timestamp. The time taken to process this step for each video lasts only a few minutes.

\subsubsection{Noise Removal}
In this section, we identify yaw drift bias as a source of noise in the annotated head orientation and discuss our strategy for its elimination. 

\bheading{Noise source.}
Gyroscope data was utilized to annotate head orientations for each frame. The gyroscope measures the angular rotational forces of the head in three axes (yaw, pitch, and roll), enabling the virtual coordinates to remain static regardless of where the user looks. This is vital for accurate head orientation tracking over time. However, the yaw component of the gyroscope data is prone to noise. Consequently, the VR coordinate system slowly drifts along the yaw axis. This form of noise is referred to as yaw drift bias \cite{foxlin1998miniature}.
The yaw drift error accumulates over the course of viewing sessions, resulting in inconsistencies in the labeled data from later parts of the camera \observations. As such, yaw drift bias significantly affects the accuracy of the annotated head poses in each camera recording.

The yaw drift can be detected by observing the inconsistency among two video frames at two different times.
Assuming $f_i, f_{i'}$ are two frames at time $t_i$ and $t_{i'}$, and $v_i, v_{i'}$ are their annotated head orientation vectors recorded from the gyroscope, respectively. Assuming the head orientations captured by $f_i$ and $f_{i'}$ are similar, e.g., the victim looks directly at the camera.
If the yaw drift occurs, there is an inconsistency between the annotated frame, \ie, $f_i - f_{i'} \neq 0$.

\bheading{Noise removal strategy.}
One method to counteract this bias involves modeling the yaw drift as a linear function of time and subtracting the yaw drift component from the annotated data. Linear modeling has been widely used for approximating data in regression and forecasting tasks \cite{weisberg2005applied}. In our case, it is possible to approximate the function parameters from two data points, $(f_i, v_i)$ and $(f_{i'}, v_{i'})$.
Let $YD = \theta{x} + \theta_0$ be the linear function modeling the yaw drift. If the component of a 3D head orientation vector $v_i$ is represented as $(v_{i,1}, v_{i,2}, v_{i,3})$, then the parameter $\theta$ and $\theta_0$ can be calculated from the head orientation vectors $v_i$ and $v_{i'}$ as:
\[\theta = (v_{i',1} - v_{i,1})/(t_{i'} - t_{i})\]
\[\theta_0 = v_{i,1}\]
Eliminating the bias involves two fundamental quaternion operations, $create\_quat(axis, deg)$ and $rotate(q, v)$ \cite{ben2014tutorial}. $create\_quat(axis, deg)$ is the function to build a rotation operation rotating a vector for $deg$ degrees around an axis specified by the unit vector $axis$. $rotate(q, v)$ is the function returning a 3D vector after applying the rotation operation $q$ on the vector $v$.
Let $[v_1, .., v_k, .., v_N]$ be the annotated head orientations of a camera recording.
For every head orientation vector $v_k$, a rotation operation $q_i = create\_quat([1, 0, 0], -YD(t_i))$ is constructed to rotate $v_k$ for $-YD(t_i)$ degrees along the yaw axis. The resulting $u_k = rotate(q_i, v_i)$ is the head orientation vector after the yaw drift bias is eliminated.

\end{document}

%% file: macro.tex
\def\Snospace~{\S{}}
\renewcommand*\sectionautorefname{\Snospace}
\def\sectionautorefname{\Snospace}
\def\subsectionautorefname{\Snospace}
\def\subsubsectionautorefname{\Snospace}
\def\chapterautorefname{\Snospace}

\newcommand{\bheading}[1]{{\vspace{2pt}\noindent{\textbf{#1}}}}
\newcommand{\iheading}[1]{{\vspace{2pt}\noindent{\textit{#1}}}}

\newcounter{note}[section]
\renewcommand{\thenote}{\thesection.\arabic{note}}
\newcommand{\zy}[1]{\refstepcounter{note}{\bf\textcolor{red}{$\ll$ZY~\thenote: {\small\sf #1}$\gg$}}}
\newcommand{\xz}[1]{\refstepcounter{note}{\bf\textcolor{blue}{$\ll$XZ~\thenote: {\small\sf #1}$\gg$}}}
\newcommand{\anh}[1]{\refstepcounter{note}{\bf\textcolor{orange}{$\ll$Anh~\thenote: {\small\sf #1}$\gg$}}}

\newcommand{\TODOxz}[1]{\refstepcounter{note}{\bf\textcolor{green}{$\ll$TODO (XZ)~\thenote: {\small\sf #1}$\gg$}}}

\newcommand{\etal}{\emph{et al.}\xspace}
\newcommand{\etc}{\emph{etc}\xspace}
\newcommand{\ie}{{i.e.}\xspace}
\newcommand{\eg}{\emph{e.g.}\xspace}

\newcommand{\figurewidth}{\columnwidth}
\newcommand{\secref}[1]{\mbox{Sec.~\ref{#1}}\xspace}
\newcommand{\secrefs}[2]{\mbox{Sec.~\ref{#1}--\ref{#2}}\xspace}
\newcommand{\figref}[1]{\mbox{Fig.~\ref{#1}}}
\newcommand{\tabref}[1]{\mbox{Table~\ref{#1}}}
\newcommand{\appref}[1]{\mbox{Appendix~\ref{#1}}}
\newcommand{\ignore}[1]{}

\newcounter{packednmbr}
\newenvironment{packedenumerate}{
\begin{list}{\thepackednmbr.}{\usecounter{packednmbr}
\setlength{\itemsep}{4pt}
\addtolength{\labelwidth}{4pt}
\setlength{\leftmargin}{25pt}
\setlength{\listparindent}{\parindent}
\setlength{\parsep}{3pt}
\setlength{\topsep}{3pt}}}{\end{list}}

\newenvironment{packeditemize}{
\begin{list}{$\bullet$}{
\setlength{\labelwidth}{6pt}
\setlength{\itemsep}{2pt}
\setlength{\leftmargin}{\labelwidth}
\addtolength{\leftmargin}{\labelsep}
\setlength{\parindent}{1pt}
\setlength{\listparindent}{\parindent}
\setlength{\parsep}{1pt}
\setlength{\topsep}{1pt}}}{\end{list}}


\newcommand{\sysname}{\mbox{\textsc{Intrude}}\xspace}
\newcommand{\detector}{\textsf{\small Detector}\xspace}


\newcommand{\observation}{\textit{recording}\xspace}
\newcommand{\observations}{\textit{{recordings}}\xspace}
\newcommand{\Observations}{{Recordings}\xspace}
\newcommand{\Observation}{{Recording}\xspace}

\newcommand{\IDTES}{\textsf{\small ID Test Set}\xspace}
\newcommand{\IDTAS}{\textsf{\small ID Training Set}\xspace}
\newcommand{\IDTRS}{\textsf{\small ID Training Set}\xspace}
\newcommand{\HMEDS}{\textsf{\small Head Pose Dataset}\xspace}

\newcommand{\RC}{\textit{RC}\xspace}
\newcommand{\RCs}{\textit{RCs}\xspace}
\newcommand{\DNNEstimator}{{head orientation estimation model}\xspace}
\newcommand{\VideoIdentifier}{{trace-fingerprint matching model}\xspace}

\newcommand{\truth}{\textit{truth}\xspace}
\newcommand{\short}{\textit{shallow}\xspace}
\newcommand{\hpose}{\textit{htrace}\xspace}
\newcommand{\std}{\textit{std}\xspace}

\newcommand{\degrees}{\degree\xspace}

%% file: 1-intro-new.tex
\section{Introduction}


Virtual Reality (VR) is a rapidly growing technology  with a projected market value of \$22 billion in 2025~\cite{vr-market}. Meta, one of the industry's leaders, has sold approximately 20 million Oculus VR headsets as of March 2023~\cite{quest-sale1,quest-sale2} and invested billions in its VR venture. Beyond the popularity in gaming and entertainment, VR head-mounted displays (HMDs) are transforming a wide range of industries, including military training, medical operation, and virtual conferencing \cite{radianti2020systematic, hsieh2018preliminary}.

The immersive nature of VR HMDs has given rise to various prevalent applications, one of which is the 360\degree~video.
360\degree~videos are widely used in areas such as virtual tours, live concerts, and immersive storytelling \cite{argyriou2020design, nassani2021showmearound}.
Unlike traditional 2D videos with limited viewing perspective on a 2D frame, 
360\degree~videos enable viewers wearing an HMD to freely explore the video content in all directions by simply moving their heads, as if they were physically present in the scene.

VR technology presents unique privacy challenges due to its collection and storage of extensive personal data~\cite{vr-privacy-1,vr-privacy-2,vr-privacy-3,vr-privacy-4}. Researchers have identified several attacks aimed at inferring VR keystrokes~\cite{luo2022holologger,al2021vr,ling2019know,meteriz2022keylogging, slocumgoing, zhang2023s,wu2023privacy}. In this paper, we unveil a new threat that identifies the titles of 360\degree~videos being viewed inside the HMD. The leakage of video titles reveals significant information about users' personal interests, opinions, and even religions \cite{xu2014watching}, which unintended third parties have exploited for pushing political agendas \cite{bork1988} or conducting blackmail \cite{scamconsumer2020}. 
Note that video identification is not necessarily related to user identification. For example, video identification attacks can target a group of users and reveal the political or religious interests of certain organizations or regions,
without performing user identification. 
%
While side-channel attacks identifying 2D videos displayed on smartphones, PCs, and TVs have been previously studied \cite{reed2016leaky,schuster2017beauty,gu2019traffic,enev2011televisions,xu2014watching}, no research has demonstrated the feasibility of side-channel attacks on 360\degree\ videos displayed on a VR HMD.

\bheading{\sysname.} We present {\sysname}, a v{\textbf{I}}deo ide{\textbf{N}}tification a\textbf{T}tack towa\textbf{R}d virt\textbf{U}al reality HM\textbf{D}s on 360\degree\ vid\textbf{E}os  by leveraging a  new \textit{contactless} side channel -- user head movements. 
While the VR content displayed inside the HMD remains inaccessible to external attackers,
we discover that the head-movement-based interaction between the user and the HMD creates a side channel that is completely exposed to the public. As video content drives the user to move or fix his/her head \cite{lookAround17}, there is a subtle
relationship between the user's head movement and the displayed 360\degree~video, which can be exploited to infer video titles without direct access to the HMD.
By taking advantage of the fact that the victim cannot see the physical world when using the HMD, the attacker can freely record the victim's head movements. After extracting the head movement from the camera recording and matching it with fingerprints of videos of interest, 
\sysname can infer the title of the playing 360\degree\ video. 

Compared to prior camera-recording-based attacks on smartphones and PCs \cite{chen2018eyetell,yue2014blind,sun2016visible}, \sysname is easier to launch and harder to detect as VR users are oblivious to their physical surroundings when wearing the HMD. 
The attackers have more flexibility in choosing when, where, and how to record the victim's head movement. 
Furthermore, since \sysname utilizes a new side-channel that does not exist in 2D video viewing, conventional countermeasures designed for 2D video identification, such as varying video encoding configuration for traffic analysis attacks and adjusting screen brightness for screen reflection attacks~\cite{schuster2017beauty,xu2014watching}, become ineffective. \looseness=-1

\bheading{Challenges and Approaches.}
Realizing \sysname\ requires addressing two new challenges in the context of 360\degree\ video identification on VR HMDs. {\it First}, it is non-trivial to extract the victim's head movement trace from the image pixels of the recording. Given that HMDs cover the majority of users' faces  and users may look around in all directions during 360\degree\ video viewing, existing head pose estimators \cite{bermejo2020eyeshopper,ruiz2018fine,patacchiola2017head,zhu2016face} harnessing facial features for limited front-facing positions cannot be directly used.
To overcome this challenge, we propose an HMD-based {\it head movement estimation} scheme to extract the full-spectrum head movement trace from the image pixels of the camera recording via a deep convolutional neural network (DCNN). 

{\it Second}, the correlation between the video title and the head movement is implicit and subtle, making it challenging to fingerprint and identify the video. Fingerprinting a video directly by head movement traces is ineffective due to the fluctuation of these traces in response to changes in human conditions or environments. Consequently, treating the head movement trace as the fingerprint may lead to poor matching with the victim's actual trace, even if they embed similar head movement patterns. Hence, we propose to fingerprint a video by saliency maps as they can capture the stable features embedded within the fluctuating head movement traces across different trials with the same video. We design a {\it trace-fingerprint matching} framework to match the extracted head movement trace with the video saliency fingerprints via a multi-modality model, enabling video identification.

\bheading{Evaluations.}  We validate \sysname through extensive evaluations that involve 31 users, 
635 videos, and two HMDs.
In the default indoor attack scenario, \sysname achieves the top-1,2,3 accuracy of 96\%,
99\%, 99\% for 360\degree~video identification. The robustness analysis under various recording environments shows that the attack maintains a stable video identification accuracy across different recording distances, lighting, backgrounds, and angles. Moreover, the risk of \sysname is validated in the {\it open-world} identification scenario. 

\bheading{Ethical Consideration.}
All experiments involving human subjects in this study have been approved by the IRB. We have only used \sysname to perform attacks on the datasets described in this paper. 
\sysname has never been used in other ways or released to other parties.

\bheading{Contributions.}
The contributions of this paper include:
\begin{packeditemize}
\item We present the first 360\degrees video identification attack on VR HMDs through a new \textit{contactless} side channel -- user head movements (\autoref{sec_mot} and \autoref{sec:overview}). This threat has not been identified and is fundamentally different from previous attacks. 
\item We build new techniques to realize the attack, i.e., a head movement estimation scheme to obtain the victim's trace (\autoref{sec:hme}) and a trace-fingerprint matching framework for video identification (\autoref{sec:identifier}). The cores of these two system components are two novel neural networks, one that can extract head orientations from recordings of 360\degrees HMD viewing and the other one that can match different input modalities for video inference. 
\item We perform an extensive evaluation (\autoref{section_setup} and \autoref{section_eva}) of the proposed
attack system \sysname, including validation of system components and parameters, baseline comparison, robustness analysis, and open-world identification attacks.
\end{packeditemize}